\documentclass[12pt]{article}
\usepackage{amssymb}
\usepackage{amsmath}
\usepackage{cite}
\usepackage{cprotect}
\usepackage{hyperref}

\usepackage{subcaption,tikz}

\input xy
\xyoption{all}
\usepackage{enumitem}

\def\sqr#1#2{{\vcenter{\vbox{\hrule height.#2pt
            \hbox{\vrule width.#2pt height#1pt \kern#1pt
                  \vrule width.#2pt}\hrule height.#2pt}}}}

\def\square
 {\mathop{\mathchoice{\sqr{12}{15}}{\sqr{9}{12}}
{\sqr{6.3}{9}}{\sqr{4.5}{9}}}}

\def\sqra#1#2#3{{\vcenter{\vbox{\hrule height.#2pt
            \hbox{\vrule width.#2pt height#1pt \kern5pt 
#3
                  \vrule width.#2pt}\hrule height.#2pt}}}}

\def\tsquare#1{\sqra{12}{15}{#1}}

\numberwithin{equation}{section}

\numberwithin{table}{section}

\begin{document}

\vspace*{-2cm}
\begin{flushright}
      \texttt{CERN-TH-2022-128}
\end{flushright}

\vspace*{.6cm}
\begin{center}

{\large\bf Decomposition, condensation defects, and fusion}

\vspace*{0.2in}

Ling Lin$^1$, Daniel G. Robbins$^2$, Eric Sharpe$^3$

\begin{tabular}{cc}
{\begin{tabular}{l}
$^1$ CERN\\
Theory Department\\
CH-1211 Geneva, Switzerland \end{tabular}}
&
{\begin{tabular}{l}
$^2$ Department of Physics\\
University at Albany\\
Albany, NY 12222 \end{tabular}}
\end{tabular} 

\begin{tabular}{c}
{\begin{tabular}{l}
$^3$ Department of Physics MC 0435\\
850 West Campus Drive\\
Virginia Tech\\
Blacksburg, VA  24061 \end{tabular}}
\end{tabular}

{\tt ling.lin@cern.ch},
{\tt dgrobbins@albany.edu},
{\tt ersharpe@vt.edu}

\end{center}

\noindent
In this paper we outline the application of decomposition
to condensation defects and their
fusion rules.
Briefly, a condensation defect is obtained by gauging a higher-form
symmetry along a submanifold, and so there is a natural interplay
with notions of decomposition, the statement that $d$-dimensional quantum field theories with
global $(d-1)$-form symmetries are equivalent to disjoint unions of
other quantum field theories.
We will also construct new (sometimes non-invertible) defects,
and compute their fusion products, again utilizing decomposition.
An important role will be played in all these analyses by
theta angles for gauged higher-form symmetries, which can be used to
select individual universes in a decomposition.

\begin{flushleft}
August 2022
\end{flushleft}

\newpage

\tableofcontents

\section{Introduction}

Decomposition \cite{Hellerman:2006zs} is now understood as the
statement that a $d$-dimensional quantum field
theory with a global $(d-1)$-form symmetry
is equivalent to a disjoint union of other $d$-dimensional quantum field
theories, known as universes 
(see e.g.~\cite{Sharpe:2022ene} for a recent review).  
Typical examples include two-dimensional gauge theories
with trivially-acting subgroups of the gauge group 
\cite{Pantev:2005rh,Pantev:2005wj,Pantev:2005zs}.
It is also known \cite{Durhuus:1993cq,Moore:2006dw} that
unitary two-dimensional topological field theories are equvialent to
disjoint unions of theories, a result which was argued
in \cite{Komargodski:2020mxz,Huang:2021zvu} to be a special case of
decomposition in the sense of \cite{Hellerman:2006zs}, utilizing
non-invertible symmetries.

Decomposition has been applied in a number of contexts, see
e.g.~\cite{Sharpe:2022ene} for a recent overview.  
In this paper, we will apply decomposition to condensation defects,
defined in \cite{Roumpedakis:2022aik,Choi:2022zal}
as follows.  Consider a $d$-dimensional quantum field theory with a global
$k$-form symmetry, and restrict
to a $(d-p)$-dimensional submanifold $\Sigma$.
Along $\Sigma$, gauge the the restriction of 
that global symmetry (assuming there is no
obstruction due to anomalies).  Along the worldvolume $\Sigma$,
this appears to be a gauged $(k-p)$-form symmetry, obtained as a 
`condensation' of the $k$-form symmetry defects on the codimension $p$ submanifold.
Such a gauging is described as a
higher gauging or as $p$-gauging the $k$-form symmetry, see also 
\cite{Kong:2014qka,Else:2017yqj,Carqueville:2017ono,Carqueville:2018sld,Hsin:2019fhf,Gaiotto:2019xmp,Mulevicius:2020bat,Johnson-Freyd:2020usu,Kong:2020cie,Kong:2020wmn,Johnson-Freyd:2020twl,Carqueville:2021dbv,Koppen:2021kry,Carqueville:2021edn}
for other discussions.
The resulting theory along $\Sigma$, obtained by gauging the
restriction of the higher-form symmetry, is a
condensation defect.
Such defects may be non-invertible under fusion, and serve as an explicit
construction of non-invertible symmetries in dimensions larger than two,
which
has seen a surge of interest very recently, see, e.g.~\cite{Heidenreich:2021xpr,Sharpe:2021srf,Nguyen:2021yld,Nguyen:2021naa,Koide:2021zxj,Choi:2021kmx,Kaidi:2021xfk,Bhardwaj:2022yxj,Hayashi:2022fkw,Arias-Tamargo:2022nlf,Choi:2022jqy,Cordova:2022ieu,Bashmakov:2022jtl,Damia:2022bcd,Choi:2022rfe,Antinucci:2022eat}.

In this paper
we outline how decomposition can be applied to condensation defects
and various analogues
and their fusion rules,
following
\cite{Roumpedakis:2022aik,Choi:2022zal,Kong:2020cie}.

We begin in section~\ref{sect:rev} with a short review of 
decomposition, focusing on examples of most direct relevance to this 
paper, namely orbifolds and topological field theories.  
We also discuss how one can recover individual universes by gauging
the higher-form symmetry (with a theta angle that distinguishes the
components).  For ordinary orbifolds in two dimensions, this gauging
was discussed in \cite{Sharpe:2019ddn}.

In~\ref{sect:review_cond_def}, we then illustrate, 
after a brief review of condensation defects and higher gauging, 
a rather simple, but direct, application of decomposition to condensation defects.
Namely, we discuss $p$-gauging the $(d-1)$-form symmetry in a decomposing theory.
This results in condensation defects that are
projectors onto universes along their worldvolumes, and in fact are
equivalent to local projection operators.
For completeness, and because they are very
much in the overall spirit of the rest of this paper,
we briefly discuss these `condensation defect projectors' formally and illustrate concrete computations 
in two-dimensional orbifolds.

In section~\ref{sect:decomp-def} we turn to a more intricate interplay between decomposition and condensation defects.
Specifically, we use decomposition to observe, in section~\ref{sect:mult}, that sometimes, fusion ring coefficients described as topological field theories
are equivalent to integer multiplicities.
For codimension-one condensation defects in 3d, these TFTs have a one-form symmetry responsible for decomposition, which only emerges as two defects collide.
In section~\ref{subsec:FusionCoefficients}
we illustrate in examples how these originate from potential one-form symmetries of the individual defects, which are obstructed from bulk-defect interactions.  In section~\ref{subsec:TPOs} we illustrate how the requisite topological point operators arise in the cases where decomposition occurs.

Finally, in section~\ref{sect:defects} we propose other defects,
which are motivated by condensation defects, but which are not themselves
condensation defects.  On a worldvolume of codimension $p$,
in a theory with a global $k=(d-p-1)$-form symmetry, these 
proposed defects are 
obtained by gauging a $k$-form symmetry along the worldvolume.
To be clear, this is not the same as $p$-gauging the $k$-form symmetry,
as that results in a gauging which, along the worldvolume,
looks like a $(k-p)$-form gauging, instead of the $k$-form symmetry gauged here.
These proposed defects are therefore not the same as
condensation defects, and need not be topological; nevertheless,
we argue that, at least formally, they appear to have similar properties,
as evidenced by e.g.~their fusion rings, which we compute in examples.
Our examples include defects in ordinary orbifolds as well as
in orbifolds by 2-groups.
As part of our analysis, we discuss gauging 2-form symmetries in
three-dimensional orbifolds \cite{Pantev:2022pbf}, extending
results of \cite{Sharpe:2019ddn} on gauging 1-form symmetries in
two-dimensional theories.

In passing, when gauging higher-form symmetries, we will use
corresponding theta angles to select particular universes from a 
decomposition.  Theta angles for gauged higher-form symmetries
in other contexts have also been discussed in
e.g.~\cite{Hsin:2020nts}.

To summarize, in this paper we will give several examples illustrating
the interplay between decomposition, condensation defects, and their
fusion products.

As this paper was nearing publication, we were informed that
related results will also appear in
\cite{Bhardwaj:2022lsg,Bartsch:2022mpm}.

\section{Decomposition and gauging higher-form symmetries}
\label{sect:rev}

In this section we will review pertinent aspects of
decomposition, which is the observation that $d$-dimensional theories
with global $(d-1)$-form symmetries are equivalent to
disjoint unions of quantum field theories.  Decomposition
has been studied in numerous examples,
see e.g.~\cite{Sharpe:2022ene} for a recent review.
In this paper, we will frequently
utilize examples in two-dimensional ordinary orbifolds
and in topological field theories, in which there are multiple
dimension-zero operators (and hence a global $(d-1)$-form symmetry),
and our review will focus on examples of this form.

\subsection{Orbifolds in two dimensions}
\label{sect:2dorb:rev}

Orbifolds in which a subgroup
of the orbifold group acts trivially\footnote{
Gauging a trivially-acting group or a noneffectively-acting
group (in which a subgroup acts trivially) may seem counterintuitive,
but was extensively studied in two-dimensional orbifolds and
gauge theories (such as abelian theories with nonminimal charges)
in e.g.~\cite{Pantev:2005rh,Pantev:2005wj,Pantev:2005zs},
which covered material ranging from existence and possible unitarity issues
to massless spectra,
mirrors, and quantum cohomology rings, and whose
conclusions formed the basis of the original work on
decomposition \cite{Hellerman:2006zs}.
The meaning of the related notion of
`trivially-acting one-form symmetries' in three dimensions
was recently discussed in
\cite{Pantev:2022kpl,Pantev:2022pbf}.
A discussion in the language of topological defect lines will appear in
\cite{rsvtoappear}.
}
are
common examples in which decomposition arises, and which we shall
utilize later in this paper.
In this subsection
we will review examples of this form, and how
the global one-form symmetry can be gauged to select out
a universe in the decomposition, results which we shall
utilize later.

Briefly, in an
orbifold $[X/\Gamma]$ in two dimensions
(meaning, a sigma model into target
$X$ with gauged $\Gamma$ action on $X$),
with
\begin{equation} \label{eq:ext}
1 \: \longrightarrow \: K \: \longrightarrow \: \Gamma \: \longrightarrow \:
G \: \longrightarrow \: 1,
\end{equation}
where $K$ acts trivially on $X$,
it was argued in \cite{Hellerman:2006zs} that
\begin{equation}  \label{eq:decomp:orb:noncentral}
{\rm QFT}\left( [X/\Gamma] \right)
\: = \: {\rm QFT}\left( \left[ \frac{ X \times \hat{K}}{G} \right]_{\hat{\omega}}
\right),
\end{equation}
where $\omega$ denotes discrete torsion, and $\hat{K}$ the set of
irreducible representations of $K$.
(See e.g.~\cite{Robbins:2020msp} 
for a generalization to the case that the orbifold
$[X/\Gamma]$ has discrete torsion.)

In the special case that $\Gamma$ is a central extension of $G$ by
$K$, so that $K$ lies within the center of $\Gamma$, the $G$ action on
$\hat{K}$ is trivial, and the expression above simplifies to
\begin{equation} \label{eq:decomp:orb:central}
{\rm QFT}\left( [X/\Gamma] \right)
\: = \: \coprod_{\rho \in \hat{K}} {\rm QFT}\left( [X/G]_{\rho(\omega)}
\right),
\end{equation}
where $\omega \in H^2(G,K)$ classifies the extension~(\ref{eq:ext})
and $\rho(\omega) \in H^2(G,U(1))$ defines discrete torsion
in the corresponding orbifold $[X/G]$.

For example, consider the orbifold $[X/D_4]$, where the center of
$D_4$, which is ${\mathbb Z}_2$, acts trivially.   
This example was studied in \cite[section 5.2]{Hellerman:2006zs}.
Since the group $D_4$ is a central extension
\begin{equation}
1 \: \longrightarrow \: {\mathbb Z}_2 \: \longrightarrow \:
D_4 \: \longrightarrow \: {\mathbb Z}_2 \times {\mathbb Z}_2 \:
\longrightarrow \: 1,
\end{equation}
we can apply decomposition in the form~(\ref{eq:decomp:orb:central}) to see
that
\begin{equation}  \label{eq:decomp:d4}
{\rm QFT}\left( [X/D_4] \right) \: = \: 
{\rm QFT}\left( [X/{\mathbb Z}_2 \times {\mathbb Z}_2] \right) \, \coprod \,
{\rm QFT}\left( [X/{\mathbb Z}_2 \times {\mathbb Z}_2]_{\rm d.t.} \right).
\end{equation}
We will apply this example to condensation defects
in e.g.~section~\ref{sect:d4}.

A non-central extension example is $[X/{\mathbb H}]$,
where ${\mathbb H}$ is the eight-element group of unit
quaternions $\{ \pm 1, \pm i, \pm j, \pm k\}$,
and $\langle i \rangle \cong {\mathbb Z}_4$ acts trivially.
This example was studied in \cite[section 5.4]{Hellerman:2006zs}.
In this case, ${\mathbb H}$ can be expressed as a non-central extension
\begin{equation}
1 \: \longrightarrow \: {\mathbb Z}_4 \: \longrightarrow \:
{\mathbb H} \: \longrightarrow \: {\mathbb Z}_2 \: \longrightarrow \: 1.
\end{equation}
Since this extension is not central, we apply decomposition in
the more general form~(\ref{eq:decomp:orb:noncentral}) to get
\begin{equation}
{\rm QFT}\left( [X/{\mathbb H}] \right) \: = \:
{\rm QFT}\left( X \right) \, \coprod \,
{\rm QFT}\left( [X/{\mathbb Z}_2] \right) \, \coprod \,
{\rm QFT}\left( [X/{\mathbb Z}_2] \right).
\end{equation}
(Of the four irreducible representations of ${\mathbb Z}_4$,
two are invariant under $G = {\mathbb Z}_2$, and
$G$ interchanges the remaining two.)
We will apply this example to condensation defects
in section~\ref{sect:h}.

In \cite{Sharpe:2019ddn}, gauging the 1-form symmetry in a decomposing
two-dimensional orbifold was described.  By picking a theta angle for
the gauging, one can select out individual universes in a decomposition.
To make this paper self-contained, we briefly outline those methods
here, as we will use such gaugings later.

For simplicity, we take the worldsheet $\Sigma = T^2$, and consider
an orbifold $[X/\Gamma]$ as above, which has a global
\footnote{
We use the notation $B^qK$ for $q$-form symmetries, as this is standard
in mathematics; other references use instead $K^{[q]}$.
} $BK = K^{[1]}$ symmetry.
We shall describe partition functions in which the $B\tilde{K}$ symmetry
is gauged.  First, recall that the partition function of a more nearly
ordinary
orbifold $[X/\Gamma]$ on worldsheet $\Sigma = T^2$ has the standard form
(see e.g.~\cite[section 8.3]{Ginsparg:1988ui})
\begin{equation}
Z_{T^2}\left( [X/\Gamma] \right) \: = \:
\frac{1}{|\Gamma|} \sum_{g, h \in \Gamma, gh=hg} 
{\scriptstyle g} \square_h,
\end{equation}
where the sum is over commuting pairs of elements $g, h \in \Gamma$, and
\begin{equation}
{\scriptstyle g} \square_h
\end{equation}
denotes the contribution to the path integral 
from maps from $\Sigma = T^2$
into $X$ with 
branch cuts along distinct cycles defined by $g, h$ (equivalently, maps from 
rectangles into $X$ such that the images of one pair of sides are
related by $g$ and the images of the other pair of sides are related
by $h$).  The fact that $K$ acts trivially simplifies this sum; the
sectors
\begin{equation}
{\scriptstyle g} \square_h
\end{equation}
map to corresponding sectors of an orbifold $[X/G]$, and the partition
function of the decomposition can be derived by simplifying the result,
as described in e.g.~\cite{Hellerman:2006zs}.

Gauging a $B\tilde{K}$ symmetry, for $\tilde{K} \subset K$, has
partition function \cite{Sharpe:2019ddn}
\begin{equation}
Z\left( \left[ [X/\Gamma] / B\tilde{K} \right] \right)
\: = \: \frac{1}{| \tilde{K} |} \sum_{z \in H^2(\Sigma,\tilde{K})
 = \tilde{K}} \epsilon(z) \left[
\frac{1}{|\Gamma|} \sum_{gh = hgz} {\scriptstyle g} \tsquare{z}_h
\right],
\end{equation}
where the sum is over $g, h \in \Gamma$ such that
$gh = hgz$, the figure
\begin{equation}
 {\scriptstyle g} \tsquare{z}_h
\end{equation}
denotes maps into $X$ with branch cuts along $g$, $h$, twisted by $z$
as above,
and $\epsilon \in {\rm Hom}(\tilde{K},U(1))$ 
is the theta angle arising in gauging
$B \tilde{K}$, which selects out the universe(s) appearing in the result.

We briefly summarize here two examples, also discussed in
\cite{Sharpe:2019ddn}.

First, consider the orbifold $[X/D_4]$, where $K = {\mathbb Z}_2 
\subset D_4$ acts trivially.  As discussed in \cite{Sharpe:2019ddn},
depending upon the choice of $\epsilon$, one finds
\begin{equation}
Z\left( \left[ [X/D_4] / B {\mathbb Z}_2 \right] \right)
\: = \:
\left\{ \begin{array}{cl}
Z\left( [X/ {\mathbb Z}_2 \times {\mathbb Z}_2] \right)
& \epsilon(-1) = +1,
\\
Z\left( [X/{\mathbb Z}_2 \times {\mathbb Z}_2]_{\rm d.t.} \right)
& \epsilon(-1) = -1,
\end{array}
\right.
\end{equation}
corresponding to the two universes in the decomposition~(\ref{eq:decomp:d4})
of $[X/D_4]$.

A second example studied in \cite{Sharpe:2019ddn}, and which we will use
later, involves the non-central extension $[X/{\mathbb H}]$ orbifold.
Here, although a ${\mathbb Z}_4 \subset {\mathbb H}$ acts trivially,
only a ${\mathbb Z}_2$ subgroup is central, and only that part corresponds
to an invertibly-realized one-form symmetry.  Gauging that $B {\mathbb Z}_2$
in the form above, from \cite{Sharpe:2019ddn} we recall\footnote{
In point of fact, the reference \cite{Sharpe:2019ddn} formally tried
to discuss gauging the $B{\mathbb Z}_4$.  However, the only possible
contributions to the partition function are from $z \in {\mathbb Z}_2
\subset {\mathbb Z}_4$, and so gauging a $B {\mathbb Z}_2$ instead
can be accomplished by just a factor of $2$.
}
\begin{equation}
Z\left( \left[ [X/{\mathbb H}] / B {\mathbb Z}_2 \right] \right)
\: = \: \left\{ \begin{array}{cl}
Z\left( [X/{\mathbb Z}_2] \, \coprod \, [X/{\mathbb Z}_2] \right)
& \epsilon(-1) = +1,
\\
Z\left( X \right) & \epsilon(-1) = -1.
\end{array}
\right.
\end{equation}

\subsection{Topological field theories}

So far we have discussed orbifolds in two-dimensional theories,
in which a subgroup of the gauge group acts trivially.
That trivially-acting subgroup is responsible for the appearance
of a global $(d-1)$-form symmetry, implemented by topological
point-like operators.

Now, any theory with such topological point-like operators should
also have a global $(d-1)$-form symmetry, possibly realized non-invertibly,
and hence decompose.  Examples that will play an important role
later in this paper include some topological field theories. 
Specifically, unitary topological
field theories with semisimple local operator algebras have
multiple dimension-zero operators, and hence decompose,
into disjoint unions of what are known as
invertible field theories, meaning theories whose Fock spaces are
one-dimensional.
This was first discussed in e.g.~\cite{Durhuus:1993cq,Moore:2006dw},
and the later works~\cite{Komargodski:2020mxz,Huang:2021zvu} observed
that this is a special case of decomposition, by virtue of presence
of non-invertible dimension-zero operators.

For example, consider
two-dimensional Dijkgraaf-Witten theory 
for a finite group $G$.  This is a unitary toppological field
theory with a semisimple local operator algebra, which by the
criteria above should decompose, and in fact there is a second
way of understanding it:  it is also an orbifold of a point.
Specifically, it is of the form $[X/G]$ where $X$ is a point and
the entire orbifold group $G$ acts trivially, and so it has a
decomposition, from our previous discussion. 
In particular, 
this theory decomposes into a disjoint union
of invertible field theories, indexed by irreducible projective\footnote{
Projective with respect to the element of $H^2(G,U(1))$ that defines
the twisting of the Dijkgraaf-Witten theory.
}
representations of
$G$ (see e.g.~\cite{Robbins:2020msp} for a more general discussion).  
For example, if $G  = {\mathbb Z}_2$, and the Dijkgraaf-Witten
theory is untwisted, the theory is equivalent to two invertible field
theories -- essentially, trivial field theories defined solely by
Euler counterterms.

To be clear, not every topological field theory so decomposes.
\begin{itemize}
\item For one, the theory must admit a local operator algebra.
Chern-Simons theories in three dimensions, unless noneffectively gauged
as in \cite{Pantev:2022kpl,Pantev:2022pbf}, 
do not admit such an algebra, and so do not decompose.
\item Even if there is a local operator algebra, we also emphasize that
we only speak of decomposition in unitary cases.  For example,
the topological subsector of the
A model with target ${\mathbb P}^n$ formally may be equivalent to a disjoint
union; however, the full quantum field theory with that target does not
decompose.  As we are interested in the full quantum field theory,
not just a topological subsector, we emphasize the importance of unitarity.
\end{itemize}

So far we have discussed orbifolds and topological
field theories, but we emphasize that results on
decomposition are not remotely restricted to these families of
examples, but in fact
have been studied much more widely in gauge theories.

\section{Condensation defect projectors}
\label{sect:review_cond_def}

In this section we will construct `condensation defect projectors,'
special cases of condensation defects obtained by $p$-gauging
a $(d-1)$-form symmetry.  We begin with a short overview of
condensation defects.

\subsection{Overview of condensation defects}

In this subsection we will briefly review the notion of condensation
defects and $p$-gauging $k$-form symmetries, following \cite{Roumpedakis:2022aik}.

Consider a $d$-dimensional system with an (invertible) $k$-form symmetry $B^k K$.
Then, $p$-gauging this symmetry on a codimension $p \leq k+1$ subspace $\Sigma$ amounts to summing over insertions of (or, `condensing') the $(d-k-1)$-dimensional topological defects, which generate the $k$-form symmetry, on all $(d-k-1)$-cycles of $\Sigma$.  The resulting defect along $\Sigma$ is known
as a condensation defect.
Formally, if we let $\eta(\gamma)$ denote a symmetry operator of $B^k K$ along
a $(d-k-1)$ cycle $\gamma$, then for compact $\Sigma$
the condensation defect along
$\Sigma$ is\footnote{
The form of the numerical factor follows from the fact
that, along $\Sigma$, one is gauging a $(k-p)$-form symmetry,
and the factors take into account various levels of
gauge transformations, gauge transformations of gauge transformations,
and so forth.)  We would like to thank S.-H.~Shao for an explanation
of this point. 
} \cite{Roumpedakis:2022aik}
\begin{equation}\label{eq:general_def_cond_def}
S_{\epsilon}(\Sigma) \: = \: 
\frac{ | H^{k-p-1}(\Sigma, K) | }{ | H^{k-p}(\Sigma, K)|}
\frac{ | H^{p-k-3}(\Sigma, K) | \cdots }{ | H^{k-p-2}(\Sigma,K)| \cdots}
\sum_{\gamma \in H_{d-k-1}(\Sigma, K)} \epsilon(\gamma) \, \eta(\gamma).
\end{equation}

In the expression above,
$\epsilon(\gamma)$ is an analogue of a theta angle for the gauging.
In general, when gauging a $k$-form symmetry $B^k K$ on a space $X$,
one can add a theta angle, determined by an element of cohomology of
the classifying space for $B^k K$, namely $B (B^k K) = B^{k+1} K$.
In the path integral, the corresponding gerbes on $X$ are equivalent
to maps $\phi: X \rightarrow B^{k+1} X$, 
so given $\omega \in H^{\rm dim X}_{\rm sing}(
B^{k+1} K, U(1))$, we can associate a phase
\begin{equation} \label{eq:theta1}
\int_X \phi^* \omega \: \in \: U(1).
\end{equation}
For example, if this is an ordinary gauge theory (meaning $k=0$),
in which case $K$ need not be abelian, then ordinary theta angles
can be understood this way.  In that case, $\omega \in H^{\rm dim X}( BK,
U(1))$ corresponds essentially to a characteristic class,
and then the phase~(\ref{eq:theta1}) is implemented as
\begin{equation}
\exp\left( i \theta \int_X {\rm Tr}\, F \wedge \cdots \wedge F \right),
\end{equation}
in the usual fashion. 

In the present case, we are $p$-gauging a $k$-form symmetry,
which along the codimension-$p$ defect $\Sigma$, is equivalent to
gauging a $(k-p)$-form symmetry.  As a result, the theta angles are
classified by elements of 
\begin{equation}
H^{d-p}_{\rm sing}\left( B^{k-p+1} K, U(1) \right).
\end{equation}
In principle, one expects that
the phases $\epsilon(\gamma)$ should then be given
by analogues of Chern-Simons forms computed using a form of descent. 
For example, if $p=k$ and $K$ is finite, 
these theta angles correspond to
elements of discrete torsion on $\Sigma$, in
\begin{equation}
H^{d-p}_{\rm sing}(BK, U(1)) \: = \: H^{d-p}_{\rm group}(K, U(1)).
\end{equation}
(Compare \cite{Dijkgraaf:1989pz,Sharpe:2000qt}.)  This was utilized
in e.g.~\cite{Roumpedakis:2022aik,Choi:2022zal}.  
However, in general theta angles will be different,
and may, for example, correspond to other modular-invariant-type phases
such as momentum/winding lattice shift factors 
\cite{Sharpe:2003cs,Cheng:2022nso}
that play an important role in asymmetric toroidal orbifolds, and also
arise from equivariant structures on tensor field potentials.

In passing, there is a map
\begin{equation}
\Omega: \: H^p( B^q K, U(1)) \: \longrightarrow \: H^{p-1}( B^{q-1} K,
U(1)),
\end{equation}
the loop space functor discussed in e.g.~\cite[section 3.3]{Pantev:2022pbf}.
In general, $\Omega$ is not an isomorphism.  For example, 
from the universal coefficients theorem, the fact that $B^k G = K(G,k)$,
and results in
\cite[appendix C]{clement}, one finds
\begin{equation}
H^4_{\rm sing}( B^2 {\mathbb Z}_4, U(1)) \: = \: {\mathbb Z}_8,
\end{equation}
which $\Omega$ maps to $H^3_{\rm sing}( B {\mathbb Z}_4, U(1)) = {\mathbb Z}_4$,
which is clearly not isomorphic.
That said,
we will see in section~\ref{sect:proj:formal} that
in the special case of matching degrees,
\begin{equation}
H^n_{\rm sing}(B^n K, U(1)) \: = \: H^1_{\rm sing}(BK, U(1)) \: = \:
{\rm Hom}(K, U(1)).
\end{equation}

Just as in an ordinary gauging procedure, which in this language would be an instance of 0-gauging, there can be obstructions in form of anomalies to $p$-gauging.
For example, the obstructions to 1-gauging a 1-form symmetries in a 3d theory would be non-trivial crossing relations between the topological line defects which generate the 1-form symmetry \cite{Roumpedakis:2022aik}.
In the following, we will always assume that such obstructions are absent when we talk about condensation defects.

When two such codimension $p$ condensation defects collide along a common worldvolume $\Sigma$, one can compute the fusion product from the fusion rules of the topological defects $\eta(\gamma)$ generating the $k$-form symmetry that has been $p$-gauged.
In general, the result is a non-invertible fusion rule \cite{Roumpedakis:2022aik},
\begin{align}\label{eq:general_fusion_rule}
      S(\Sigma) \times S'(\Sigma) = \sum_i c_i S_i(\Sigma) \, .
\end{align}

In concrete examples, it may also be possible to give a `microscopic' description of condensation defects, in terms of a Lagrangian field theory on $\Sigma$ coupled to the $d$-dimensional bulk --- an approach that we will utilize in Section \ref{sect:decomp-def}.
In such cases, the fusion coefficients $c_i$ in \eqref{eq:general_fusion_rule} 
can sometimes be described as topological field theories \cite{Roumpedakis:2022aik}.
However, as also noted implicitly in \cite[footnote 3]{Choi:2022zal}, 
sometimes those topological field theory coefficients are
equivalent to numbers, a simple multiplicity.
We will discuss this in greater detail in section~\ref{sect:decomp-def}.

\subsection{Condensation defect projectors}
\label{sect:proj}

In this section we will discuss `condensation defect projectors,'
which are defined to be the condensation defects arising from $p$-gauging
a $(d-1)$-form symmetry in a $d$-dimensional quantum field theory.
We will see that the result has a simple universal form.

In a $d$-dimensional quantum field theory with a global
$(d-1)$-form symmetry, the corresponding symmetry generators are pointlike,
and their linear combinations can be used to build
projectors, as we shall discuss shortly.
(This is one reason why such a quantum field theory decomposes into
distinct universes, and gauging the $(d-1)$-form symmetry projects onto
one of the universes.)

An important consequence of the existence of these projectors is that the $(d-1)$-form symmetry is 0-gaugeable, i.e., there is no obstruction to summing over insertions of those pointlike operators.
Therefore, it is also $p$-gaugeable for any $p>0$ \cite{Roumpedakis:2022aik}, which produces condensation defects associated to submanifolds $\Sigma$ of any dimension $(d-p)$.
In fact, along $\Sigma$, the higher gauging is 
equivalent to gauging a $(d-1)-p = (d-p-1)$-form symmetry,
which undoes the decomposition along $\Sigma$.
Put another way, this is equivalent to
the insertion of a projection operator for one of the 
universes of the ambient theory on $\Sigma$.
As a result, this is a defect which is invisible to one universe (the one projected onto),
but appears as an insertion of zero to every other universe.
When $\Sigma$ is real codimension one (a domain wall),
this is effectively akin to a bandpass filter.

Now, condensation defects can sometimes be simplified\footnote{
We would like to thank Y.~Choi for a useful discussion of this fact.
}.
For example, if one $p$-gauges a global $(d-q)$-form symmetry, 
so that the symmetry generators live on submanifolds of dimension $q-1>0$,
and the submanifold $\Sigma$ is $q$-connected, then one expects
that condensation defects $S(\Sigma)$ on $\Sigma$ are trivial,
as $H_{q-1}(\Sigma,K) = 0$.

In the present case, the condensation defects we will construct
(for $p$-gauging a $(d-1)$-form symmetry) will be equivalent to
operators on a collection of points, as many points as the number of connected
components of $\Sigma$ 
(with an operator on one point in each component of $\Sigma$),
corresponding to elements of $H_0(\Sigma,K)$.  
As a result, condensation defects corresponding to $p$-gauged
global $(d-1)$-form symmetries will be equivalent to 
(collections of) local projection operators, for any $p$.

Nevertheless, we will find it instructive to quickly step through the
details and perform some consistency tests.

\subsubsection{Formal construction}
\label{sect:proj:formal}

Formally, to $p$-gauge a $(d-1)$-form symmetry $B^{d-1} K$ along $\Sigma$
defines a condensation defect, according to \eqref{eq:general_def_cond_def}, of the form
\begin{equation}
S_{R}(\Sigma) \: = \: 
\frac{ | H^{d-p-2}(\Sigma,K)| \, | H^{d-p-4}(\Sigma,K) | \cdots }{
| H^{d-p-1}(\Sigma,K) | \, | H^{d-p-3}(\Sigma,K) | \cdots}
 \sum_{\gamma \in
H_0(\Sigma,K) = K} \epsilon_R(\gamma) \, p(\gamma),
\end{equation}
where 
$\epsilon_R \in \hat{K} = {\rm Hom}(K,U(1))$ is a theta angle
for the symmetry gauging, corresponding to universe $R$.
It is straightforward to check that 
\begin{equation}
\frac{ | H^{d-p-2}(\Sigma,K)| \, | H^{d-p-4}(\Sigma,K) | \cdots }{
| H^{d-p-1}(\Sigma,K) | \, | H^{d-p-3}(\Sigma,K) | \cdots}
\: = \: \frac{ |K|^{\pm \chi} }{ |H^{d-p}(\Sigma,K)|},
\end{equation}
so as $\Sigma$ has dimension $d-p$, if we assume it is compact and
connected, then up to Euler counterterms, we have that
\begin{equation}
S_R(\Sigma) \: = \: 
\frac{1}{|K|} \sum_{\gamma \in
 K} \epsilon_R(\gamma) \, p(\gamma).
\end{equation}

Now, let us describe the theta angles $\epsilon$ more explicitly.
As discussed previously, theta angles $\epsilon$ appearing when $p$-gauging
a $(d-1)$-form symmetry are classified by elements of
\begin{equation}
H^{d-p}_{\rm sing}\left( B^{(d-1) - p +1} K, U(1) \right)
\: = \: H^{d-p}_{\rm sing}\left( B^{d-p} K, U(1) \right).
\end{equation}
Now, from the universal coefficients theorem and the fact that
$B^{d-p} K = K(K,d-p)$ has no homology in nonzero degree less than $d-p$,
\begin{eqnarray}
H^{d-p}_{\rm sing}\left( B^{d-p} K, U(1) \right)
& = &
{\rm Hom}\left( H_{d-p}( B^{d-p}K ), U(1) \right),
\\
& = & {\rm Hom}\left( K, U(1) \right),
\end{eqnarray}
using the Hurewicz theorem to compute
\begin{equation}
H_{d-p}\left( B^{d-p} K \right) \: = \:
H_{d-p}\left( K(K, d-p) \right) \: = \: 
\pi_{d-p}\left( K(K,d-p) \right) \: = \: K.
\end{equation}
See also \cite{stackexchange} for further discussion of this result.
In any event, we see that when $p$-gauging a $(d-1)$-form symmetry $B^{d-1} K$,
the possible theta angles $\epsilon$ 
are classified by elements of Hom$(K,U(1))$, for any $p$.

The operators $p(\gamma)$ are pointlike topological operators that generate
the global $(d-1)$-form symmetry in the $d$-dimensional theory -- they are
the operators which, in some sense, are responsible for the decomposition
of the theory.  
Assuming, for simplicity, that the $(d-1)$-form symmetry is realized invertibly with finite and abelian $K$, they obey
\begin{equation}
p(\gamma) \, p(\lambda) \: = \: p(\gamma \lambda)
\end{equation}
for all $\gamma, \lambda \in K$.

The projection onto the universe associated with an irreducible representation $R$ of $K$\footnote{
The observation that universes
are associated with irreducible representations of $K$, and not
representations of a higher-form
analogue such as $B^p K$ for some $p$, was discussed in
\cite[appendix B]{Pantev:2022kpl}.
} is implemented by the local operator\footnote{
Ultimately this is a consequence of Wedderburn's theorem in mathematics.
In two-dimensional theories, projectors for more general cases were
given in \cite[section 2.2]{Sharpe:2021srf}.  The fact that universes
are associated with irreducible representations of $K$, and not a higher-form
analogue such as $B^p K$ for some $p$, was discussed in
\cite[appendix B]{Pantev:2022kpl}.
}
\begin{equation}\label{eq:projection_op}
\Pi_{R} \: = \: \frac{1}{|K|} \sum_{\gamma \in K}
\chi_R\left( \gamma^{-1} \right) \, p(\gamma),
\end{equation}
where $\chi_R$ is the character associated to $R$.
The sum corresponds to gauging $B^{d-1}K$ in the full spacetime, which `undoes' the decomposition \cite{Sharpe:2019ddn}.
For example, in a Lagrangian description, the effect of each $p(\gamma)$ is
to twist the theory by a (higher) $K$-gerbe with characteristic class $\gamma$,
and summing over those (higher) gerbes implements the projection, precisely as in \cite{Sharpe:2019ddn}.

As a result, for the $R$th universe, if we identify the theta angle with the coefficients appearing in \eqref{eq:projection_op},
\begin{equation}
\epsilon_R(\gamma) \: = \: \chi_R\left(\gamma^{-1}\right),
\end{equation}
then we can write very simply
\begin{equation}
S_R(\Sigma) \: = \: \Pi_R |_{\Sigma},
\end{equation}
which, as claimed, establishes the condensation defect associated to irreducible representation $R$ as insertions of the projector operator
$\Pi_R$ along the defect $\Sigma$.
(In fact, as noted earlier, $S_R(\Sigma)$ is equivalent to an insertion
of a local projection operator $\Pi_R$ at a point in each connected
component of $\Sigma$.)

For observers in universe $R$, an insertion of $S_R(\Sigma)$ anywhere
is effectively invisible.  However, correlation functions between
operators in any other universe different from $R$ will vanish if the defect
$S_R(\Sigma)$ is inserted along any submanifold $\Sigma$, just as inserting
a local projection operator will annihilate correlation functions in different
universes.  (Decomposing theories do not obey cluster decomposition
\cite{Hellerman:2006zs}, and as the condensation defect is topological,
we cannot avoid this conclusion simply by moving the defect
$\Sigma$ far away from observables.)

Fusion rules are now easy to compute.
In principle, they follow immediately from the basic property of
projectors:
\begin{equation}
\Pi_R \Pi_S \: = \: \delta_{R,S} \Pi_R.
\end{equation}
In this context, we can repeat this directly from the definition of
the defect above.  To avoid Euler counterterms, let us work on
a defect $\Sigma = T^2$.

Then, we compute
\begin{eqnarray}
S_R(\Sigma) \times S_S(\Sigma)
& = &
\frac{1}{|K|^2} \sum_{\gamma, \lambda \in K} 
\chi_R\left( \gamma^{-1} \right)
\chi_S\left( \lambda^{-1} \right)
p(\gamma) p(\lambda),
\\
& = &
\frac{1}{|K|^2} \sum_{\gamma, \lambda \in K} \chi_S\left( \lambda^{-1} \gamma
\right) \chi_R\left( \gamma^{-1} \right) 
p( \lambda ),
\\
& = &
\frac{\delta_{R,S}}{|K|} \sum_{\lambda \in K} \chi_R\left( \lambda^{-1} \right)
p(\lambda) \: = \:
\delta_{R,S} \, S_R(\Sigma),
\end{eqnarray}
using the identity (see e.g.~\cite[appendix B]{Sharpe:2021srf})
\begin{equation}
\frac{1}{|G|} \sum_{g \in G} \chi_R(ag) \chi_S(g^{-1}b) \: = \:
\frac{\delta_{R,S}}{\dim R} \chi_R(ab)
\end{equation}
for $G$ a finite group and $R, S$ irreducible representations of $G$.
This also is exactly as expected from the fact that the
condensation defect $S_R(\Sigma)$ is equivalent to a collection
of local operators $\Pi_R$, one at a point of each connected component
of $\Sigma$.

\subsubsection{Orbifolds in $d=2$}

Next, to be completely thorough, 
let us make this more explicit in two dimensions in a concrete family of
examples.
Consider an orbifold $[X/\Gamma]$, where
\begin{equation}
1 \: \longrightarrow \: K \: \longrightarrow \: \Gamma \: \longrightarrow
\: G \: \longrightarrow \: 1
\end{equation}
is a central extension of the finite group $G$ by another finite (and abelian)
group $K$.  Assume that $K$ acts trivially on $X$, so that the orbifold
has a global one-form symmetry, and so decomposes.

Now, let us imagine computing the partition function of a two-dimensional
theory on worldsheet $\Sigma$ with a condensation defect inserted along
a line $L$, corresponding to 1-gauging the global one-form symmetry
$BK$ (with theta angle $\epsilon_R$, corresponding to universe $R$ in the
decomposition of $[X/\Gamma]$).  In principle, the partition function
of the orbifold $[X/\Gamma]$ itself is 
a sum over contributions from the constituent universes.  Inserting
a condensation defect projector along $L$ should project out the contributions
from all but one of those universes, as we shall see explicitly.

In the spirit of \cite[section 6]{Roumpedakis:2022aik},
if we break up the worldsheet $\Sigma$ into regions to the left and
right of the line $L$, and imagine orbifolds over each of those
regions independently (to the extent that the global geometry allows),
we are led to a partition function which, for $\Sigma = T^2$ for
simplicity, has the form
\begin{eqnarray}
Z\left( \Sigma, L \right) 
& = &
\frac{1}{|K|} \sum_{z \in K} \epsilon_R(z)
\left[
\frac{1}{| \Gamma |^2} \sum_{g, h_1, h_2 \in \Gamma}
\raisebox{-10pt}{
\begin{tikzpicture}
\draw (0,0) -- (2,0);
\draw (0,0) -- (0,0.75);
\draw (0,0.75) -- (2,0.75);
\draw (2,0) -- (2,0.75);
\draw[dashed] (1,0) -- (1,0.75);
\node [left] at (0,0.4) {$g$};
\node [above] at (0.5,0.75) {$h_1$};
\node [above] at (1.5,0.75) {$h_2$};
\node [right] at (1,0.4) {$L$};
\end{tikzpicture}
} \right],
\label{eq:z-sigma-l-1}
\\
& = &
\frac{1}{|K|} \sum_{z \in K} \epsilon_R(z)
\left[ \frac{1}{| \Gamma |^2} \sum_{g, h_1, h_2 \in \Gamma}
{\scriptstyle g} \tsquare{z}_{h_1 h_2^{-1}} \right].
\end{eqnarray}
In effect, we break the $T^2$ into a pair of $T^2$'s, joined along
$L$, with the orbifold partition function of one $T^2$ corresponding
to a sum over $g, h_1$, and that of the other $T^2$ corresponding to
a sum over $g, h_2$, where $h_1, h_2$ are split at the location of
the defect $L$.  Graphically, if we identify the defect $L$ with an 
edge of holonomy $v$, then
from demanding that both squares close, and that one is twisted by $z \in K$
as in \cite{Sharpe:2019ddn} and section~\ref{sect:2dorb:rev}, we have
the two conditions
\begin{equation}
g h_1 v h_1^{-1} \: = \: 1, \: \: \:
gz h_2 v h_2^{-1} \: = \: 1,
\end{equation}
and eliminating $v$ implies
\begin{equation}
g h_1 h_2^{-1} \: = \: h_1 h_2^{-1} g z.
\end{equation}
(The reader could also reach this conclusion by inspection of the
diagram in~(\ref{eq:z-sigma-l-1}).)

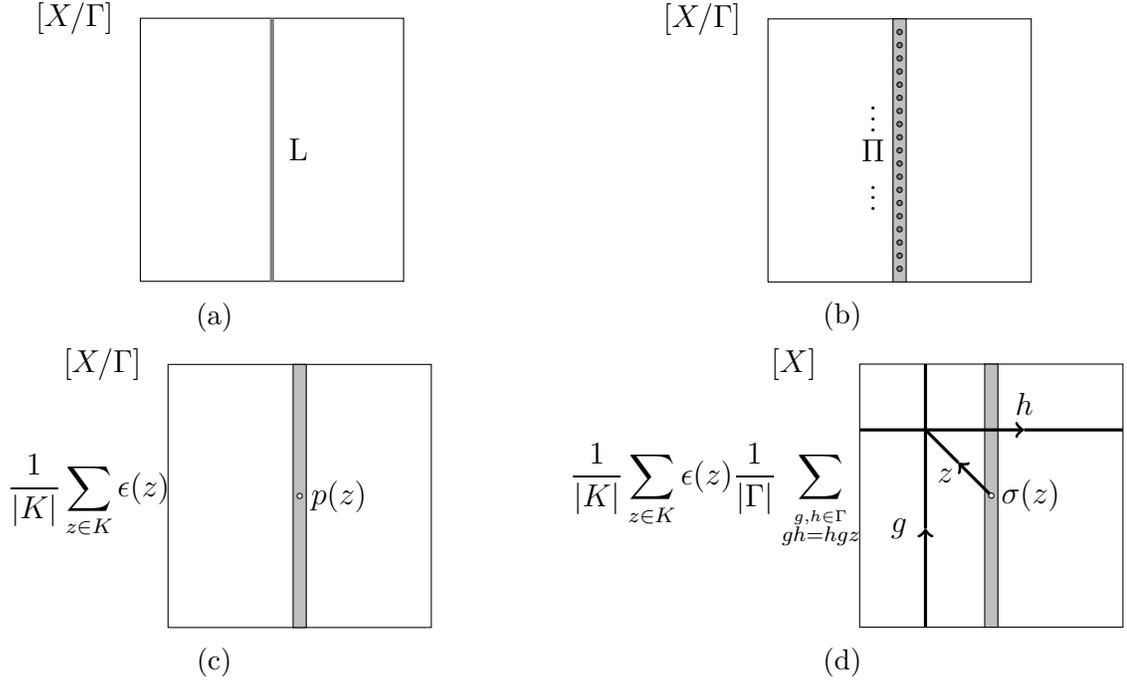
\begin{figure}
\begin{subfigure}{0.5\textwidth}
\centering
\begin{tikzpicture}[scale=1.75]
   \draw (0,0) -- (2,0) -- (2,2) -- (0,2) -- (0,0);
   \node at (-0.5,2) {$[X/\Gamma]$};
   \draw[ultra thick,gray] (1,0) -- (1,2);
   \node at (1.2,1) {L};
\end{tikzpicture}
\caption{}
\end{subfigure}
\begin{subfigure}{0.5\textwidth}
\centering
\begin{tikzpicture}[scale=1.75]
   \draw (0,0) -- (2,0) -- (2,2) -- (0,2) -- (0,0);
   \node at (-0.5,2) {$[X/\Gamma]$};
   \draw[fill=lightgray] (0.95,0) rectangle (1.05,2);
   \foreach \y in {1,...,19} {\draw[fill=gray] (1,0.1*\y) circle (0.02);}
   \node at (0.8,1) {$\Pi$};
   \node at (0.8,1.3) {$\vdots$};
   \node at (0.8,0.7) {$\vdots$};
\end{tikzpicture}
\caption{}
\end{subfigure}
\begin{subfigure}{0.5\textwidth}
\centering
\begin{tikzpicture}[scale=1.75]
   \draw (0,0) -- (2,0) -- (2,2) -- (0,2) -- (0,0);
   \node at (-0.5,2) {$[X/\Gamma]$};
   \draw[fill=lightgray] (0.95,0) rectangle (1.05,2);
   \draw[fill=white] (1,1) circle (0.02);
   \node at (1.3,1) {$p(z)$};
   \node at (-0.6,1) {$\displaystyle{\frac{1}{|K|}\sum_{z\in K}\epsilon(z)}$};
\end{tikzpicture}
\caption{}
\end{subfigure}
\begin{subfigure}{0.5\textwidth}
\centering
\begin{tikzpicture}[scale=1.75]
   \draw (0,0) -- (2,0) -- (2,2) -- (0,2) -- (0,0);
   \node at (-0.5,2) {$[X]$};
   \draw[fill=lightgray] (0.95,0) rectangle (1.05,2);
   \draw[very thick,->] (0.5,0) -- (0.5,0.75);
   \draw[very thick] (0.5,0.75) -- (0.5,2);
   \node at (0.3,0.75) {$g$};
   \draw[very thick,->] (0,1.5) -- (1.25,1.5);
   \draw[very thick] (1.25,1.5) -- (2,1.5);
   \node at (1.25,1.7) {$h$};
   \draw[very thick,->] (1,1) -- (0.75,1.25);
   \draw[very thick] (0.75,1.25) -- (0.5,1.5);
   \node at (0.65,1.15) {$z$};
   \draw[fill=white] (1,1) circle (0.02);
   \node at (1.3,1) {$\sigma(z)$};
   \node at (-1.1,1) {$\displaystyle{\frac{1}{|K|}\sum_{z\in K}\epsilon(z)\frac{1}{|\Gamma|}\sum_{\stackrel{g,h\in\Gamma}{gh=hgz}}}$};
\end{tikzpicture}
\caption{}
\end{subfigure}
\caption{We illustrate the computation of a $T^2$ correlation function in the orbifold theory $[X/\Gamma]$ with one-cycle $L$ wrapped by a condensation defect, as shown in (a).  In (b)-(d) the cycle is shaded gray for visualization purposes, but the only insertions along the cycle are the point operators shown.  The prescription is to insert a projection point operator $\Pi$ at every vertex point in some sufficiently fine triangulation of $L$, as in (b).  Since $L$ is connected, it is sufficient to insert $\Pi$ at a single point, and we can write $\Pi$ as a sum over twist fields $p(z)$, shown in (c).  Finally, to compute these correlation functions in terms of the parent theory $[X]$, we lift each diagram and sum over all consistent ways of inserting $\Gamma$ lines.  In the lift, the $p(z)$ operator becomes an operator $\sigma(z)$ sitting on the end of a $z$ line.  We can choose where that line joins the other lines.  One choice (where they all meet at a junction of degree five) is shown in (d) and it is implicit in our lift that the product of lines around the junction should be the identity, giving the requirement $gh=hgz$.}
\label{fig:1g1f2d}
\end{figure}

Now, write $h = h_1 h_2^{-1}$.  One of those two group elements
$h_1, h_2$ is now redundant,
and summing over its values gemerates a factor of $| \Gamma |$.  
This implies
\begin{eqnarray}
Z\left( \Sigma, L \right) & = &
\frac{1}{|K|} \sum_{z \in K} \epsilon_R(z)
\left[ \frac{1}{ | \Gamma | } \sum_{g,h \in \Gamma} 
{\scriptstyle g} \tsquare{z}_{h} \right],
\\
& = & 
Z\left( \Sigma, \left[ [X/\Gamma] / B K \right] \right),
\\
& = & Z\left( \mbox{universe }R \right)
\end{eqnarray}
using the description of gauged one-form symmetries
in orbifolds in \cite{Sharpe:2019ddn}, as reviewed
in section~\ref{sect:2dorb:rev}.
An illustration of this defect and the conclusion above
is in figure~\ref{fig:1g1f2d}.

Thus, we see that, formally, the partition function of the
two-dimensional orbifold $[X/\Gamma]$ with 
a condensation defect projector along $L$ is equivalent to 
the partition function of the $BK$-gauged orbifold, which is
the same as that of the universe corresponding to $R$.

\section{Decomposition in fusion coefficients}
\label{sect:decomp-def}

\subsection{Formal discussion}
\label{sect:mult} 

While the coefficients appearing in the fusion rules \eqref{eq:general_fusion_rule} of general condensation defects are TFTs, many examples discussed in \cite{Roumpedakis:2022aik} end up with coefficients that do have an interpretation as a simple multiplicity.

The fundamental observation here is that in such examples, the unitary topological
field theories all come with semisimple local operator algebras.
As such, they decompose into disjoint unions of invertible field theories, as reviewed earlier.
For example, in fusions of condensation defects of 3d Maxwell theory, Chern-Simons theory, or discrete gauge theories, the coefficients $c_i$ in \eqref{eq:general_fusion_rule} are all themselves 2d $\mathbb{Z}_n$ gauge theories for appropriate $n$.
Such a theory decomposes into $n$ isomorphic universes \cite{Hellerman:2006zs,Hellerman:2010fv}, and hence, gives $n$ identical copies of $S_i$.

That said, as noted earlier, not every topological field
theory decomposes.  In particular, the Chern-Simons theories 
appearing as topological-field-theory coefficients in
\cite{Choi:2022zal} are typically not equivalent to integers.

We should also clarify that even when the topological field theory
decomposes, it still contains slightly more information than just an
integer, in the form of Euler counterterms.  As counterterms, they can be
shifted, but for some applications their canonical values may be
pertinent.  We give here two examples of those counterterms.

First, for two-dimensional
untwisted Dijkgraaf-Witten theory for a finite group $G$,
the partition function on a connected Riemann
surface of genus $g$ is
\begin{equation}
Z_g(G) \: = \: \sum_R \left( \frac{ \dim R }{|G|} \right)^{2-2g},
\end{equation}
where the sum is over (untwisted) irreducible representations $R$ of
$G$.  This form precisely reflects the decomposition:
the universes into which two-dimensional Dijkgraaf-Witten theory
decomposes are indexed by the irreducible representations $R$,
and one can read off the Euler counterterms in universe $R$,
given by
\begin{equation}
\ln\left(  \frac{ \dim R }{|G|} \right).
\end{equation}

Second, consider the $G/G$ model at level $k$, for $G$ connected and
simply-connected.
Here, the partition function equals the dimension of the corresponding
Chern-Simons Hilbert space
(see e.g.~\cite[section 3.4]{Blau:1993hj}),
which at genus $g$ is
\cite[equ'n (3.15)]{Verlinde:1988sn}, \cite{gukovprivate},
\cite[equ'n (C.4)]{Komargodski:2020mxz}
\begin{equation}
Z_g \: = \: \sum_{i} \left( S_{0 i} \right)^{2-2g},
\end{equation}
where $S_{0i}$ is proportional to the quantum dimension of the
integrable representation $i$, and the sum is over integrable
representations of the Kac-Moody algebra at level $k$.

In the following, we will elaborate on such multiplicities in 
examples of condensation defects of three-dimensional theories, and provide a `microscopic' explanation for the appearance of a decomposing TFT fusion coefficient.
Namely, we will exhibit the emergence of a one-form symmetry on the worldvolume as two defects fuse, which can be understood as a cancellation of obstructions to have a one-form symmetry on each individual defect.

In examples with Lagrangian descriptions, as discussed in \cite[section 6]{Roumpedakis:2022aik}, the individual condensation defects have a 2d $BF$-type worldvolume action,
\begin{align}
\label{eq:BFAction}
      \frac{i n}{2\pi} \int_\Sigma \phi \, dA
\end{align}
As mentioned abve, these 2d $\mathbb{Z}_n$ gauge theories decompose by themselves.
When coupled to the 3d bulk, the one-form symmetry is broken; but when bringing two defects close, there is a linear combination of the two individual one-form symmetries that is unbroken.

\subsection{Fusion coefficients in ${\mathbb Z}_2$ gauge theories}
\label{subsec:FusionCoefficients}

To illustrate the above story, let us take a closer look at the fusion process of condensation defects in 3d pure $\mathbb{Z}_2$ gauge theories, which itself has a $BF$-type Lagrangian,
\begin{align}
      \frac{2i}{2 \pi} \int A \, d \tilde{A} \, ,
\end{align}
with $A$ and $\tilde{A}$ two $U(1)$ gauge fields.

Including a single condensation defect $S_e$ obtained from 1-gauging the electric $\mathbb{Z}_2$ 1-form symmetry on the codimension-1 surface $\Sigma = \{x=0\}$, the total system is described by the action \cite[section 6.3.4]{Roumpedakis:2022aik}:
\begin{align}\label{eq:3d_Z2_total_action}
\begin{split}
      & \frac{2i}{2\pi} \int_{x<0} A_L \, d \tilde{A}_L + \frac{2i}{2\pi} \int_{x>0} A_R \, d \tilde{A}_R - \frac{2i}{2\pi} \int_{x=0} \Phi \, d(\tilde{A}_L - \tilde{A}_R) \, , \\
      & A_L \Big|_{x=0} = A_R \Big|_{x=0} = d\Phi \, .
\end{split}
\end{align}

Naively, the worldvolume term,
\begin{align}
      - \frac{2i}{2\pi} \int_{x=0} \Phi \, d(\tilde{A}_L - \tilde{A}_R) \equiv - \frac{2i}{2\pi} \int_{x=0} \Phi \, d\tilde{A}_{\rm{d}},
\end{align}
describes a 2d $\mathbb{Z}_2$ gauge theory, and should, by itself, decompose.
Including the background field $B \in H^2(\Sigma, \mathbb{Z}_2)$ for the one-form symmetry responsible for the decomposition, the 2d action takes the form
\begin{align}
       \frac{i}{2\pi} \int_{x=0} \Phi \, (2 d\tilde{A}_{\rm{d}} - B) \, ,
\end{align}
where $\tilde{A}_d=\tilde{A}_L-\tilde{A}_R$.

In general, a 2d $BF$-theory \eqref{eq:BFAction} with $\mathbb{Z}_n$ gauge symmetry has a $\mathbb{Z}_n$ 1-form symmetry which is generated by topological point operators $:e^{ik\phi}:$, where $k$ is an integer and $k\sim k+n$ (see for instance Appendix B of~\cite{Hellerman:2010fv}).
These operators are topological, i.e.~do not depend on the position of insertion, because the $A$ equation of motion implies that $\phi$ is constant (at least locally; on disconnected spacetimes it can in principle take different values on each component).  Moreover, in order for the action to be well-defined under large gauge transformations of $A$, even in the presence of a boundary, we require $n\phi$ to be in $2\pi\mathbb{Z}$.  Combined with the $2\pi$ periodicity of $\phi$, this explains why $k\sim k+n$, and why we have only $n$ distinct topological point operators.

However, in contrast to ordinary $BF$ theory, the scalar $\Phi$ on the condensation defect is related to
the restriction of the bulk gauge fields $A_L$ and $A_R$,
as noted in~(\ref{eq:3d_Z2_total_action}).
In particular, as gauge transformations in the bulk must still be allowed, 
\begin{align}
      & A_L \sim A_L + d \alpha_L \, , \quad A_R \sim A_R + d\alpha_R \, , \quad \Phi \sim \Phi + \alpha \, , \quad \left(\alpha = \alpha_L \big|_{x=0} = \alpha_R \big|_{x=0}\right).
\end{align}
In the presence of a non-trivial background field $B$ for the 2d 1-form symmetry, such a gauge transformation would lead to a non-integer shift of the action \eqref{eq:3d_Z2_total_action}, and, thus, to an ambiguity for the partition function.
As the bulk gauge symmetries must remain intact, this ambiguity poses an obstruction which effectively breaks the 1-form symmetry of the 2d $BF$-theory, and prevents the condensation defect from decomposing into simpler pieces.
Put another way, in terms of the topological
local operators $: \exp(i k \Phi) :$, here the coupling to the bulk means that
$\Phi$ is not gauge-invariant, and so those local operators
are not well-defined.

Interestingly, given two condensation defects, each with a $BF$-worldvolume theory, there is a partial cancellation between the obstructions for the 1-form symmetries on each defect, in the limit where they collide. 
To see this, we first start with the two defects separated by distance $\epsilon$,
\begin{align}\label{eq:Z2_defects_with_B_fields}
      & \frac{2i}{2\pi} \int_{x<0} A_L \, d \tilde{A}_L + \frac{2i}{2\pi} \int_{0<x<\epsilon} A_I \, d \tilde{A}_I  + \frac{2i}{2\pi} \int_{x>\epsilon} A_R \, d \tilde{A}_R \\
      - & \frac{i}{2\pi} \int_{x=0} \Phi_1 \, [2 d(\tilde{A}_L - \tilde{A}_I) - B_1] - \frac{i}{2\pi} \int_{x=\epsilon} \Phi_2 \, [2d(\tilde{A}_I - \tilde{A}_R) - B_2] \, ,
\end{align}
where for the purpose of illustration, we have added the 1-form symmetry backgrounds on each defect even though they must be set to zero for consistency.
The gauge symmetries of the system include the variations
\begin{align}
\begin{split}
      & \delta \Phi_1 = \alpha_1 \, , \quad \delta \Phi_2 = \alpha_2 \, , \quad \delta A_I = d\alpha_I \, , \\
      \text{with} \qquad & \alpha_1 = \alpha_I \big|_{x=0} \, , \quad \alpha_2 = \alpha_I \big|_{x=\epsilon} \, .
\end{split}
\end{align}
At finite $\epsilon$, $\alpha_{1}$ and $\alpha_2$ are independent, and each pose the obstruction to turning on non-vanishing $B_{1/2}$.
However, as $\epsilon \rightarrow 0$, we have the gauge variations $\delta \Phi_1 = \delta \Phi_2 = \alpha_I$, so that $\Phi = \Phi_1 - \Phi_2$ is gauge invariant.
Performing the analogous rearrangement of \eqref{eq:Z2_defects_with_B_fields} as in \cite{Roumpedakis:2022aik}, but including the $B$-fields, we obtain the worldvolume Lagrangian of the fused defect,
\begin{align}
-\frac{i}{2\pi}\int_{x=0}\left[ -2(\Phi_1 - \Phi_2) d(\tilde{A}_I-\tilde{A}_L) + \Phi_1 B_1  + \Phi_2 B_2 +\Phi_2\left( d\tilde{A}_L-d\tilde{A}_R\right)\right] \, .
\end{align}
Now we see that, though generic values of $(B_1, B_2)$ are still not allowed, we can correlate the 1-form symmetries of the two individual defects prior to fusion, by setting $B_1 = -B_2 = B$, in which case the worldvolume Lagrangian becomes
\begin{align}\label{eq:Z2_fusion_final_form}
-\frac{i}{2\pi}\int_{x=0}\left[ -\Phi \, [ 2d(\tilde{A}_I-\tilde{A}_L) - B] +\Phi_2\left( d\tilde{A}_L-d\tilde{A}_R\right)\right] \, .
\end{align}
Because $\Phi = \Phi_1 - \Phi_2$ is invariant under any gauge symmetries of the system, there is no obstruction as above to turning on non-trivial $B$.

This is of course just the same conclusion as the observation, that in the fusion rule \cite{Roumpedakis:2022aik},
\begin{align}
      S_e \times S_e = ({\cal Z}_{2}) S_e \, ,
\end{align}
the coefficient is a 2d $\mathbb{Z}_2$ gauge theory which does undergo decomposition, by the existence of a 1-form symmetry on the worldvolume with background field $B$.
From the discussions of this subsection, we see that this 1-form symmetry is the (anti-)diagonal subgroup of the product of two 1-form symmetries from separate defects, which by themselves are broken, but give rise to an unbroken one in the limit as the two defects fuse.

A similar story applies also to 3d Chern-Simons theory with level $2N$.
There, the condensation defects $S_n$, with $n | N$, 
constructed in \cite[section 6.2]{Roumpedakis:2022aik} 
also admit a Lagrangian description that has a $BF$-type term, namely
\begin{equation}
\frac{i n}{2\pi} \int_{x=0} d\phi (A_L - A_R).
\end{equation}
However, gauge invariance of the bulk system require the presence of an additional term, 
\begin{equation}
\frac{iN}{2\pi} \int_{x=0} A_L A_R,
\end{equation} 
which does not admit a 1-form symmetry on the worldvolume $\{x=0\}$.
Consequently, there are no `simpler' pieces into which these condensation defects decompose.
However, given two such condensation defects, an unobstructed 1-form symmetry emerges in the limit where these collide, resulting in a decomposable TFT sector.
This is again the fusion coefficient, $S_n \times S_n = {\cal Z}_n S_n$, which admits a simple interpretation as a multiplicity.

\subsection{Topological point operators in fusion coefficients}
\label{subsec:TPOs}

It is worthwhile to refine the discussion by taking a closer look at the local, or point operators, on the individual condensation defects and in their fusion.
These can be used to build projection operators onto the different universes of the decomposition associated to a global $(d-1)$-form symmetry, provided they are not bound to topological defect lines.

In the 2d $BF$-theory \eqref{eq:BFAction}, the objects charged under these point operators are Wilson loops, $\exp(i\ell\oint A)$.  When a Wilson line encircles one of our topological point operators (TPOs) $e^{i k \phi}$, it picks up a phase relative to the configuration where the TPO is outside the loop.  (For the projection operators, this means that one projection operator will transform into another when it crosses a Wilson line, so the Wilson lines have an interpretation as interfaces between different universes.)  We can also interpret the Wilson lines as the topological lines which generate the global $\mathbb{Z}_n$ 0-form symmetry, which the 2d theory \eqref{eq:BFAction} possesses.

Now consider instead a condensation defect obtained by 1-gauging a 1-form symmetry in a 3d theory.  
Though the Lagrangian description superficially takes the form of 2d $BF$ theory coupled to a 3d bulk, the story can change in a subtle, but important way.
To be concrete, let us consider the condensation defects in 3d free Maxwell theory~\cite{Roumpedakis:2022aik}.  This 3d theory has action
\begin{equation}
S=\frac{1}{g^2}\int F\wedge\star F.
\end{equation}
Since the classical equation of motion is simply $d\star F=0$, we can introduce a dual scalar $\sigma$ by (the factor of $i$ comes from the Euclidean signature)
\begin{equation}
\frac{4\pi i}{g^2}\star F=d\sigma.
\end{equation}
The Dirac quantization of $F$ implies that $\sigma\sim\sigma+2\pi$.  This theory has point operators ('t Hooft monopoles) $e^{i\alpha\sigma}$.  When $\alpha\in\mathbb{Z}$, this is a good local operator.  When $\alpha\notin\mathbb{Z}$ such operators can still make sense when they are attached to the endpoint of an electric 1-form symmetry line $\exp(\tfrac{4\pi i\alpha}{g^2}\int_\gamma\star F)$.  However, none of these point operators are topological, since $d\sigma\ne 0$ in general.  There are TPOs in the theory, but they are not free; they only occur at junctions of topological line operators.  For instance, a collection of $K$ electric symmetry lines, $\exp(\tfrac{4\pi i\alpha_j}{g^2}\int\star F)$, $j=1,\cdots,K$, can meet at a topological junction provided that $\sum_j\alpha_j\in\mathbb{Z}$.

We can create a condensation defect $S_N$ by 1-gauging a $\mathbb{Z}_N$ subgroup of the $U(1)$ electric 1-form symmetry.  For the case of a defect on a surface $x=0$, we can write an action~\cite{Roumpedakis:2022aik}
\begin{equation}
\label{eq:MaxwellDefectAction}
S=\frac{1}{g^2}\int_{x<0}F_L\wedge\star F_L+\frac{1}{g^2}\int_{x>0}F_R\wedge\star F_R+\frac{iN}{2\pi}\int_{x=0}\phi\left(dA_L-dA_R\right).
\end{equation}
By taking the $A_L$ and $A_R$ equations of motion and integrating over an infinitesimal interval in $x$ around the defect\footnote{Explicitly we can rewrite~(\ref{eq:MaxwellDefectAction}) as
\begin{equation}
S=\int\left[\frac{1}{g^2}F_L\wedge\star F_L\Theta(-x)+\frac{1}{g^2}F_R\wedge\star F_R\Theta(x)+\frac{iN}{2\pi}\phi\left(dA_L-dA_R\right)\wedge\delta(x)dx\right],
\end{equation}
where $\Theta(x)$ is the Heaviside function.  This results in an equation of motion for $A_L$
\begin{equation}
0=\frac{2}{g^2}d\star F\Theta(-x)+\frac{2}{g^2}\star F\wedge\delta(x)dx+\frac{iN}{2\pi}d\phi\wedge\delta(x)dx.
\end{equation}
} we learn that
\begin{equation}
\frac{4\pi i}{g^2}\left.\star F\right|_{x=0}=Nd\phi\quad\Rightarrow\quad \left.d\sigma\right|_{x=0}=Nd\phi.
\end{equation}
The 2d theory on the defect admits monopole operators $e^{ik\phi}$.  If $N|k$, then this is equivalent to a bulk monopole operator and the operator can leave the defect, but for $k$ not zero modulo $N$ this operator is bound to the defect.  However these are not topological, and we do not have decomposition on the single defect $S_N$.  To get a TPO we again need a nontrivial junction.  We can for instance have a bulk $\mathbb{Z}_N$ electric symmetry line $\exp(\tfrac{4\pi ik}{g^2N}\int\star F)$ which ends on the defect (hooking on to the network of lines from the 1-gauging), but no free TPOs.

On the other hand, consider the fusion of two such defects $S_N$ and $S_{N'}$.  We have the possibility of an electric symmetry line that starts on one defect and ends on the other, connecting topologically to both networks.  This will only work if the line is both $\mathbb{Z}_N$ valued {\it{and}} $\mathbb{Z}_{N'}$ valued.  In other words it must have the form $\exp(\tfrac{4\pi ik}{g^2\operatorname{gcd}(N,N')}\int\star F)$ and we have $\operatorname{gcd}(N,N')$ such configurations in total.  In the limit where we take the two defects to be coincident, these become a pair of topological point operators connected by a line along the defect.  Shrinking the line and taking the ends to be coincident produces a TPO which is bound to the defect but otherwise free.  The theory on the fusion project thus does have a $\mathbb{Z}_{\operatorname{gcd}(N,N')}$ 1-form symmetry on the defect, resulting in decomposition.  Of course this is just the decomposition of the TFT coefficient $\mathcal{Z}_{\operatorname{gcd}(N,N')}$ in the fusion
\begin{equation}
S_N\times S_{N'}=\left(\mathcal{Z}_{\operatorname{gcd}(N,N')}\right)\,S_{\operatorname{lcm}(N,N')}.
\end{equation}
We could gauge this 1-form symmetry on the defect, essentially inserting a projector built out of the TPOs, and project onto a single $S_{\operatorname{lcm}(N,N')}$ defect.

Similar considerations apply in other theories.  For instance, in $U(1)_{2N}$ Chern-Simons theory there is a non-1-anomalous $\mathbb{Z}_N$ 1-form symmetry generated by the Wilson line $\eta:=\exp(2i\int A)$, and for any divisor $n$ of $N$ we can 1-gauge a $\mathbb{Z}_n$ subgroup generated by $\eta^m$, where $N=nm$, to create a condensation defect $S_n$.  In the bulk theory the only TPOs are at junctions of Wilson lines $\exp(ia_j\int A)$ such that $a_j\in\mathbb{Z}$ and $\sum_ja_j\equiv 0\ (\mathrm{mod\ }2N)$.  Again the $S_n$ defect has no free TPOs on it, and hence no 1-form symmetry and no decomposition.  We can once more look for the possibility of a bulk symmetry line ending on the defect at a topological point.  In order to attach on to the network of lines from the 1-gauging, the bulk line must be of the form $\exp(2ikm\int A)$, i.e. it must be $(\eta^m)^k$ for some integer $k$.  However, now there is an additional twist relative to the Maxwell case.  Since these Wilson lines have non-trivial braiding, an $\eta^m$ line along the defect can produce a phase if it encircles the point where the bulk line meets the defect.  When we 1-gauge we sum over all such configurations and the phases will cancel out unless the meeting point is invariant (essentially we project onto $\mathbb{Z}_n$-invariant bulk lines), and this will happen if and only if the bulk  line has the form $\eta^{n\ell}$ for some integer $\ell$.  In summary then, the only way for a bulk line $\eta^a$ to meet the defect topologically is if $m|a$ and $n|a$, where $N=nm$.  Now if we want to fuse a pair of defects $S_n$ and $S_{n'}$, with $N=nm=n'm'$, then there can be a bulk line $\eta^a$ between them if and only if $a$ is a multiple of $n$, of $m$, of $n'$, and of $m'$.  There will be $g:=\operatorname{gcd}(n,m,n',m')$ such configurations in total, generating a $\mathbb{Z}_g$ 1-form symmetry on the fusion product (again corresponding to a $\mathcal{Z}_g$ TFT coefficient).  This matches the results in~\cite{Kapustin:2010if,Kapustin:2010hk,Roumpedakis:2022aik}.

\section{Proposal for additional defects}
\label{sect:defects}

So far in this paper we have discussed condensation defects.
In this section, we will propose a related set of objects,
which are not, to our knowledge, condensation defects,
but which in some respects seem analogous.

Let us outline our proposed defects formally.
Suppose a $d$-dimensional quantum field theory has a $k$-form symmetry,
and restrict
to a $(d-p)$-dimensional submanifold $\Sigma$.
(To be clear, when we speak of restricting to $\Sigma$,
we imagine working locally on $\Sigma \times {\mathbb R}^p$, and 
dimensionally-reducing to $\Sigma$.)
If $k = d-p-1$, then the restriction of the theory to $\Sigma$ will 
decompose (as the restriction to $\Sigma$
is a $(d-p)$-dimensional theory with a 
$(d-p-1)$-form symmetry).

Now, given any quantum field theory in $d$ dimensions, schematically with
action
\begin{equation}
S_0 \: = \: 
\int_X d^d x \, {\cal L},
\end{equation}
we are free to introduce new fields that propagate
along a submanifolds $\Sigma \subset X$.
Specifically, in our proposed defects, we introduce 
tensor field potentials and couplings along $\Sigma$ which
gauge the $k$-form symmetry on $\Sigma$,
giving a total theory with action of the form
\begin{equation}
S \: = \: S_0 \: + \: \frac{1}{g^2} \int_{\Sigma} d^{d-p} x \,
{\cal L}_1,
\end{equation}
where $1/g^2$ is related to the tension of the defect,
and ${\cal L}_1$ is the Lagrangian density for fields along the defect.

Along $\Sigma$, the result of this gauging is to project to
a subset of the universes of
the decomposition along $\Sigma$, as in \cite{Sharpe:2019ddn}.  To define the
gauging we pick a theta
angle, whose choice determines which subset is projected onto by the
gauging.  The resulting theories define our proposal for a class
of defects.

In this section, we will make that proposal explicit in examples,
both for these defects themselves as well as for
their fusion products, computed in a limit in which the defect is
massive, to minimize interactions with bulk fields.

Although the construction will be analogous to a higher gauging,
the result will not be precisely the same as a condensation defect.
For example, in a condensation defect along a codimension $p$
submanifold $\Sigma$, $p$-gauging a $k$-form symmetry of the ambient
theory looks, along $\Sigma$, like gauging a $(k-p)$-form symmetry.
By contrast, in this section we consider gauging a $k$-form symmetry
along $\Sigma$, not a $(k-p)$-form symmetry.

Since these defects are not condensation defects, they
need not be topological, for example.
Nevertheless, we believe they may be of interest,
so we define them and propose computations of fusion rules.

\subsection{Two-dimensional defects in orbifolds}

\subsubsection{Construction}

Consider a three- or four-dimensional\footnote{
The dimension indicated is that of the space on which the quantum field
theory lives, and is not related to the dimension of the target space $X$.
Also, if $X$ is not flat, then the orbifold theory should be understood
as a low-energy effective field theory, as also discussed in \cite{Pantev:2022kpl}.
} orbifold $[X/\Gamma]$,
where
\begin{equation}
1 \: \longrightarrow \: K \: \longrightarrow \: \Gamma \: \longrightarrow \:
G \: \longrightarrow \: 1,
\end{equation}
and $K$ acts trivially.  Since $K$ acts trivially, the  
theory has a $BK$ symmetry -- but not a decomposition, for which we
would need a two- or three-form symmetry, depending upon dimension.  

Now, restrict the theory to a 2-submanifold $\Sigma$.  
The restriction is a two-dimensional orbifold with a trivially-acting
subgroup, hence again a one-form symmetry, and now, a decomposition.
We can produce an analogue of a 
condensation defect by gauging that global one-form symmetry along $\Sigma$,
which, following \cite{Sharpe:2019ddn} and as reviewed
in section~\ref{sect:2dorb:rev}, 
selects out a universe (depending upon the
theta angle chosen).

So, for each 2-submanifold $\Sigma$, we now have a collection of
defects, one for each universe in the decomposition of the two-dimensional
orbifold $[X/\Gamma]$.

Now, let us compute fusion rules.  Following \cite{Sharpe:2019ddn},
the defect is obtained by gauging a 1-form symmetry $BK$
on a theory on the two-dimensional space $\Sigma$, which means the
path integral
\begin{itemize}
\item sums over $K$ gerbes, and then,
\item for each $K$ gerbe, sums over $K$-twisted bundles and maps into $X$.
\end{itemize}
In principle, in the path integral of the fusion of two defects
along the same submanifold $\Sigma$, one would like to 
tensor together the $K$ gerbes
and the $K$-twisted bundles.  

We consider these two issues in turn.  First, consider the gerbes.
Since $K$ is abelian, $K$ is a product of cyclic groups, so for simplicity
of presentation, and without loss of generality, let us suppose that $K$
is cyclic, and imagine computing a fusion product of two such defects.
Suppose one defect is defined by gauging $B {\mathbb Z}_p$,
and the other by gauging $B {\mathbb Z}_k$,
where both ${\mathbb Z}_k, {\mathbb Z}_{\ell} \subset K$.
Let us first consider the gerbes in the path integral.
In the collision, one has a product of a ${\mathbb Z}_p$ gerbe and a
${\mathbb Z}_k$ gerbe.  This product is a ${\mathbb Z}_{pk}$ gerbe,
induced from a ${\mathbb Z}_{ {\rm lcm}(p,k)}$ gerbe.
Note that ${\mathbb Z}_{pk}$ is not necessarily a subgroup of $K$,
so we cannot consistently extend the gerbes on either side to
${\mathbb Z}_{pk}$, as the groups are assumed to lie within $K$.
However,
it will always be the case that
${\mathbb Z}_{ {\rm lcm}(p,k) } \subset K$,
so we extend the gerbes on either side to ${\mathbb Z}_{ {\rm lcm}(p,k) }$
gerbes.  Doing that change of variables in the path integral
on $\Sigma \times I$
will leave the $B {\mathbb Z}_{ {\rm gcd}(p,k) }$ uncoupled.

More formally\footnote{
We would like to thank T.~Pantev for a discussion of these products.
},
\begin{equation} \label{eq:gcd-lcm}
1 \: \longrightarrow \: {\mathbb Z}_{ {\rm gcd}(p,k) } \:
\longrightarrow \: {\mathbb Z}_p \times {\mathbb Z}_k \: 
\longrightarrow \: {\mathbb Z}_{ {\rm lcm}(p,k) } \: 
\longrightarrow \: 1,
\end{equation}
which induces
\begin{equation}
H^2(\Sigma, {\mathbb Z}_{ {\rm gcd}(p,k) } ) \: \longrightarrow \:
H^2( \Sigma, {\mathbb Z}_p \times {\mathbb Z}_k) \:
\longrightarrow \: H^2( \Sigma, {\mathbb Z}_{ {\rm lcm}(p,k) } )
\: \longrightarrow \: 0.
\end{equation}
For our purposes, this means that the product of ${\mathbb Z}_p$ and
${\mathbb Z}_k$ gerbes can be described as 
${\mathbb Z}_{ {\rm lcm}(p,k) } \subset K$ gerbes, and the mapping to
${\mathbb Z}_{ {\rm lcm}(p,k) }$ gerbes has, as fiber,
${\mathbb Z}_{ {\rm gcd}(p,k) }$ gerbes.

Now that we have a picture of how to combine the gerbes,
we turn to the bundles.
Unfortunately, if $G$ is not abelian, we do not know of a well-defined way to
tensor two $G$ bundles to get another $G$ bundle.  To make sense of
this product, we borrow a trick from OPE computations of anomalies
(see e.g.~\cite[section 19.1]{Peskin:1995ev}), 
where one repairs gauge invariance by connecting two operators, separated
by a finite distance, by a Wilson line.
Here, we
extend $\Sigma$ to a box, $\Sigma \times I$,
for $I$ an interval, with the two defects at either end of the interval,
where we shrink the interval to zero size at the end of the computation.
The path integral sums over isomorphisms between the boundaries.
For bundles, this means the path integral sums over gauge fields in the
interior, generating parallel transporters which explicitly identify
boundary fields.  For gerbes, the gerbes on the ends are forced to be
isomorphic, and so we can identify them, following the gcd/lcm prescription
described above.

Let us make this more explicit.  To compute a fusion product,
we
extend $\Sigma$ to a box, $\Sigma \times I$,
for $I$ an interval, with the two defects at either end of the interval.
After doing the computation, we then shrink the interval to zero size.
For example, if $\Sigma = T^2$, we consider a box $T^2 \times I$ 
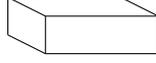
\begin{center}
\begin{tikzpicture}
\draw (0,0) -- (0,0.5);  \draw (0,0) -- (0.5,-0.25);
\draw (0,0.5) -- (0.5,0.25);  \draw (0.5,0.25) -- (0.5,-0.25);
\draw (0,0.5) -- (1.5,0.5);  \draw (0.5,0.25) -- (2,0.25);
\draw (0.5,-0.25) -- (2,-0.25);
\draw (1.5,0.5) -- (2,0.25); \draw (2,0.25) -- (2,-0.25);
\end{tikzpicture}
\end{center}
with
\begin{equation}
{\scriptstyle g_1} \tsquare{y}_{h_1},
\: \: \:
{\scriptstyle g_2} \tsquare{z}_{h_2}
\end{equation}
at either end, where we now think of $y, z \in {\mathbb Z}_{ {\rm lcm}(p,k) }$.

Since $I$ is contractible, these twisted bundles
must be isomorphic, and the edges parallel to the interval provide
parallel transporters relating the holonomies around the edges.
In particular, for this box to be nonzero requires
$y = z$ and
\begin{equation} \label{eq:conj}
g_1 \: = \: \gamma g_2 \gamma^{-1},
\: \: \:
h_1 \: = \: \gamma h_2 \gamma^{-1},
\end{equation}
for some $\gamma \in \Gamma$ (corresponding to the edges parallel to $I$).
To contribute to a twisted sector, consistency requires
\begin{equation} \label{eq:edge-comm}
g_1h_1 \: = \: h_1 g_1 y, \: \: \:
g_2h_2 \: = \: h_2g_2 z,
\end{equation}
and it is straightforward to check that so long as (\ref{eq:conj}) holds,
and $K$ is in the center of $\Gamma$, the two conditions~(\ref{eq:edge-comm})
are equivalent to one another.

Now, let us assemble these pieces.
In a limit of large mass (e.g.~$g^2 \rightarrow 0$),
the partition function for a single 
defect, gauging $BK$, on $\Sigma = T^2$ is of the form
\cite[equ'n (6.9)]{Sharpe:2019ddn}
\begin{equation}
Z \: = \: \frac{1}{|K|} \frac{1}{|\Gamma|} 
\sum_{z \in K} \sum_{gh = hgz} \epsilon_{\ell}(z) \,
{\scriptstyle g} \tsquare{z}_h,
\end{equation}
where
\begin{equation}
{\scriptstyle g} \tsquare{z}_h
\end{equation}
denotes a twisted sector of the $\Gamma$ orbifold which has been
twisted by a $K$ gerbe with characteristic class
$z \in H^2(\Sigma, K) = K$,
and $\epsilon_{\ell}(z)$ is a theta angle for the gauged one-form symmetry.
(The choice of $\epsilon_{\ell}$ determines which universe, or collection
of universes, in the decomposition 
is selected by the one-form-symmetry gauging.)

In the same limit,
the partition function of the fusion product of two such defects on
$\Sigma = T^2$,
one with a gauged ${\mathbb Z}_p$, the other with a gauged ${\mathbb Z}_k$,
is then of the form
\begin{equation}  \label{eq:orb-product}
{\rm gcd}(p,k) \frac{1}{| {\mathbb Z}_{ {\rm lcm}(p,k) } | }
 \frac{1}{|\Gamma|^2}
\sum_{z \in {\mathbb Z}_{ {\rm lcm}(p,k)}} \sum_{g_1h_1 = h_1g_1z}
\sum_{g_2h_2 = h_2g_2z}
\sum_{\gamma \in \Gamma}
 \epsilon_{\ell_1}(z)
\epsilon_{\ell_2}(z) 
\raisebox{-25pt}{
\begin{tikzpicture}
\draw (0,0) -- (0,0.5);  \draw (0,0) -- (0.5,-0.25);
\draw (0,0.5) -- (0.5,0.25);  \draw (0.5,0.25) -- (0.5,-0.25);
\draw (0,0.5) -- (1.5,0.5);  \draw (0.5,0.25) -- (2,0.25);
\draw (0.5,-0.25) -- (2,-0.25);
\draw (1.5,0.5) -- (2,0.25); \draw (2,0.25) -- (2,-0.25);
\node [below] at (1.25,-0.25) {$\gamma$};
\end{tikzpicture}
}
\end{equation}
where we sum over $g_1, h_1, g_2, h_2, \gamma \in \Gamma$ such that
\begin{equation}
g_i h_i \: = \: h_i g_i z, \: \: \:
g_1 \: = \: \gamma g_2 \gamma^{-1}, \: \: \:
h_1 \: = \: \gamma h_2 \gamma^{-1}.
\end{equation}
(The overall factor of the gcd reflects the uncoupled 
${\mathbb Z}_{ {\rm gcd}(p,k) }$ gerbe in the path integral, 
and the fact that the denominator
has a factor of the order of ${\mathbb Z}_{ {\rm lcm}(p,k) }$ reflects
the fact that the separate ${\mathbb Z}_p$ and ${\mathbb Z}_k$ gerbes
have been replaced by a gerbe of order lcm$(p,k)$.)
We shall see in examples that in general this is a rather complicated
combinatorial condition.

In the next several subsections we will work through details of examples
of these computations.  We will begin in sections~\ref{sect:z2}, \ref{sect:zp}
with a pair of relatively simple
examples, orbifolds in which the entire orbifold group acts
trivially, closely analogous to examples in \cite{Roumpedakis:2022aik}.  
Our later examples in 
sections~\ref{sect:d4}, \ref{sect:h} 
discuss more general cases, involving nonabelian
orbifolds in which only a subgroup acts trivially on the space.

\subsubsection{Example: ${\mathbb Z}_2$ gauge theory}
\label{sect:z2}

Consider the case that the three-dimensional orbifold is $[X/{\mathbb Z}_2]$
with the ${\mathbb Z}_2$ itself acting trivially, closely analogous to
examples in
 \cite[section 6.3]{Roumpedakis:2022aik} and described
earlier in section~\ref{subsec:FusionCoefficients}.  (That said,
we emphasize again that in this section we are gauging a 1-form symmetry
along the defect, not a 0-form symmetry, so this is not the same
as the condensation defects studied there.)  This theory has
a $B {\mathbb Z}_2$ symmetry, and its restriction to a two-dimensional
$\Sigma$ therefore decomposes, in this case to two identical copies of
a sigma model on
$X$.  If we gauge $B {\mathbb Z}_2$ along $\Sigma$, then depending upon
the choice of discrete theta angle, we will recover each of those two
sigma models.

As before, suppose that $\Sigma = T^2$, so that in the large tension
limit the partition function
for $S_k(\Sigma)$ is
\cite[equ'n (6.9)]{Sharpe:2019ddn}
\begin{equation}
Z \: = \: \frac{1}{|{\mathbb Z}_2|^2} 
\sum_{z \in {\mathbb Z}_2} \sum_{gh = hgz} \epsilon_k(z) \,
{\scriptstyle g} \tsquare{z}_h.
\end{equation}
Here, however, there are no contributions when $z \neq 1$, as all of the
group elements are abelian, and as $\epsilon_k(1) = +1$ for all $k$,
this reduces to
\begin{equation}
Z\left( S_k(\Sigma) \right) \: = \: \frac{1}{|{\mathbb Z}_2|^2} 
\sum_{gh = hg} {\scriptstyle g} \square_h
\: = \: Z(\Sigma, X),
\end{equation}
for both values of $k$.

Now, let us compute the fusion product~(\ref{eq:orb-product}):
\begin{equation}
(2) \frac{1}{| {\mathbb Z}_2|} \frac{1}{| {\mathbb Z}_2|^2}
\sum_{z \in {\mathbb Z}_2}  \sum_{g_1h_1 = h_1g_1z}
\sum_{g_2h_2 = h_2g_2z}
\sum_{\gamma \in \Gamma}
 \epsilon_{\ell_1}(z)
\epsilon_{\ell_2}(z) 
\raisebox{-25pt}{
\begin{tikzpicture}
\draw (0,0) -- (0,0.5);  \draw (0,0) -- (0.5,-0.25);
\draw (0,0.5) -- (0.5,0.25);  \draw (0.5,0.25) -- (0.5,-0.25);
\draw (0,0.5) -- (1.5,0.5);  \draw (0.5,0.25) -- (2,0.25);
\draw (0.5,-0.25) -- (2,-0.25);
\draw (1.5,0.5) -- (2,0.25); \draw (2,0.25) -- (2,-0.25);
\node [below] at (1.25,-0.25) {$\gamma$};
\end{tikzpicture}
}
\end{equation}
where $K = \Gamma = {\mathbb Z}_2$ in this case.
Now, since $\Gamma = {\mathbb Z}_2$ is abelian, the equation
\begin{equation}
g_i h_i \: = \: h_i g_i z
\end{equation}
can only be solved when $z=1$, and also since $\Gamma = {\mathbb Z}_2$
is abelian,
\begin{equation}
g_1 \: = \: \gamma g_2 \gamma^{-1} \: = \: g_2,
\: \: \:
h_1 \: = \: \gamma h_2 \gamma^{-1} \: = \: h_2,
\end{equation}
independent of $\Gamma$.  Thus, $g_2$ and $h_2$ are uniquely
determined by $g_1$ and $h_1$, and the sum over $\gamma$ just
contributes an overall factor of $| \Gamma | = 2$.
Thus, we find
\begin{eqnarray}
\lefteqn{
(2) \frac{1}{| {\mathbb Z}_2|} \frac{1}{| {\mathbb Z}_2|^2}
\sum_{z \in {\mathbb Z}_2}  \sum_{g_1h_1 = h_1g_1z}
\sum_{g_2h_2 = h_2g_2z}
\sum_{\gamma \in \Gamma}
 \epsilon_{\ell_1}(z)
\epsilon_{\ell_2}(z) 
\raisebox{-25pt}{
\begin{tikzpicture}
\draw (0,0) -- (0,0.5);  \draw (0,0) -- (0.5,-0.25);
\draw (0,0.5) -- (0.5,0.25);  \draw (0.5,0.25) -- (0.5,-0.25);
\draw (0,0.5) -- (1.5,0.5);  \draw (0.5,0.25) -- (2,0.25);
\draw (0.5,-0.25) -- (2,-0.25);
\draw (1.5,0.5) -- (2,0.25); \draw (2,0.25) -- (2,-0.25);
\node [below] at (1.25,-0.25) {$\gamma$};
\end{tikzpicture}
}
} \nonumber \\
& = &
(2) \frac{ 1 }{ | {\mathbb Z}_2 |^2 } \sum_{g_i, h_i} \epsilon_{\ell_1 + \ell_2}(z=1)
\, {\scriptstyle g_1} \square_{h_1},
\\
& = &
(2)  Z(X)
\end{eqnarray}
for all $\ell_{1,2}$, as $\epsilon_{\ell}(+1) = +1$ and, since
the ${\mathbb Z}_2$ acts trivially,
\begin{equation}
{\scriptstyle g} \square_h \: = \: {\scriptstyle 1} \square_1
\end{equation}
for all (commuting) pairs $g, h \in \Gamma$.

In terms of the fusion product, we interpret
the factor of 2 to mean that two copies of the defect
appear.  In other words, our two defects
$S_0(\Sigma) \cong S_1(\Sigma)$, and if we write $S_e$ for either,
the partition function above implies
\begin{equation}
S_e \times S_e \: = \: 2 \, S_e.
\end{equation}

This fusion product for closely analogous defects was also computed in
\cite[section 6.3]{Roumpedakis:2022aik}.
There, one single condensation defect $S_e(\Sigma)$ was discussed,
which here appears as a pair of defects $S_{0,1}(\Sigma)$,
and the fusion product computed there 
\cite[equ'n (6.64)]{Roumpedakis:2022aik}
matches the result above, modulo describing a disjoint union of
two copies of $S_e$ as a TFT coupled to $S_e$
(as discussed earlier in section~\ref{sect:mult}).

In sections~\ref{sect:d4}, \ref{sect:h} we will discuss more general
examples in which a $B {\mathbb Z}_2$ is gauged in an orbifold with
a trivially-acting ${\mathbb Z}_2$, and in those examples, the
two defects $S_{0,1}(\Sigma)$ will no longer be isomorphic,
though we will find that they still obey a variation of the fusion product
above.

\subsubsection{Example: $[X/D_4]$}
\label{sect:d4}

Consider an orbifold $[X/D_4]$, where $D_4$ is the eight-element
dihedral group, and the ${\mathbb Z}_2$ center of $D_4$ acts trivially
on $X$.  The resulting three-dimensional orbifold has a global
$B {\mathbb Z}_2$ symmetry.
Let $\Sigma$ be a 2-submanifold, and restrict the orbifold to $\Sigma$.
The restriction to $\Sigma$ decomposes:
\begin{equation}
[X/D_4] |_{\Sigma} \: = \: 
[X/{\mathbb Z}_2 \times {\mathbb Z}_2]|_{\Sigma} \, \coprod \, 
[X/{\mathbb Z}_2 \times {\mathbb Z}_2]_{\rm d.t.} |_{\Sigma},
\end{equation}
where the d.t.~subscript indicates discrete torsion,
as has been discussed in e.g.~\cite[section 5.2]{Hellerman:2006zs}.
If we gauge the $B {\mathbb Z}_2$ along $\Sigma$, 
as discussed in \cite[section 6.2]{Sharpe:2019ddn} and reviewed
in section~\ref{sect:2dorb:rev},
then depending upon theta angles,
we can get either $[X/{\mathbb Z}_2 \times {\mathbb Z}_2]$ or
$[X/{\mathbb Z}_2 \times {\mathbb Z}_2]_{\rm d.t.}$.
Let $S_0(\Sigma)$ denote the ${\mathbb Z}_2 \times {\mathbb Z}_2$ orbifold
without discrete torsion, and $S_1(\Sigma)$ the orbifold with
discrete torsion:
\begin{equation}
S_0(\Sigma) \: = \: [X/{\mathbb Z}_2 \times {\mathbb Z}_2] |_{\Sigma},
\: \: \:
S_1(\Sigma) \: = \: [X/{\mathbb Z}_2 \times {\mathbb Z}_2]_{\rm d.t.} |_{\Sigma}.
\end{equation}

Now, let us consider their fusion products.
To be explicit, suppose that $\Sigma = T^2$.
Then, in the limit of large mass,
the partition function for a single defect
$S_{k}(\Sigma)$ is
\cite[equ'n (6.9)]{Sharpe:2019ddn}
\begin{equation}
Z \: = \: \frac{1}{|{\mathbb Z}_2|} \frac{1}{|D_4|} 
\sum_{z \in {\mathbb Z}_2} \sum_{gh = hgz} \epsilon_k(z) \,
{\scriptstyle g} \tsquare{z}_h,
\end{equation}
where
\begin{equation}
{\scriptstyle g} \tsquare{z}_h
\end{equation}
denotes a twisted sector of the $D_4$ orbifold which has been
twisted by a ${\mathbb Z}_2$ gerbe with characteristic class
$z \in H^2(\Sigma, {\mathbb Z}_2) = {\mathbb Z}_2$,
and
\begin{equation}
\epsilon_k(z) \: = \: 
z^k \: = \:
\left\{ \begin{array}{cl}
+1 & z = 1 \mbox{ or } k=0,
\\
-1 & z = -1 \mbox{ and } k=1.
\end{array} \right.
\end{equation}

To compute the fusion, let us enumerate twisted sectors.
Following the same notation as \cite{Hellerman:2006zs,Sharpe:2019ddn},
write 
\begin{equation}
D_4 \: = \: \{ 1, z, a, b, az, bz, ab, ba = abz \},
\end{equation}
where $a^2 = 1 = z^2$, $b^2 =z$ generates the ${\mathbb Z}_2$ center, 
which is quotiented
to form ${\mathbb Z}_2 \times {\mathbb Z}_2$.  Also,
write ${\mathbb Z}_2 \times {\mathbb Z}_2 = \langle
\overline{a}, \overline{b} \rangle$, where the projection of $a, az \in D_4$
is $\overline{a} \in {\mathbb Z}_2 \times {\mathbb Z}_2$,
and the projection of $b, bz \in D_4$ is 
$\overline{b} \in {\mathbb Z}_2 \times {\mathbb Z}_2$.
To help keep track of computations,
let $A$ denote all of the $D_4$ twisted sectors appearing when
$z = +1$, and $B$ denote all of the $D_4$ twisted sectors appearing for
$z = -1$.  As discussed in \cite{Sharpe:2019ddn}, for $z=+1$,
the $D_4$ twisted sectors correspond to ${\mathbb Z}_2 \times {\mathbb Z}_2$
twisted sectors that lift to $D_4$, which almost all do, except for
sectors of the form
\begin{equation}
{\scriptstyle \overline{a}} \square_{\overline{b}}, \: \: \:
{\scriptstyle \overline{a}} \square_{\overline{ab}}, \: \: \:
{\scriptstyle \overline{b}} \square_{\overline{ab}}.
\end{equation}
The set above defines $B$.

Now, in the large mass limit,
the partition function of the fusion product is of the form~(\ref{eq:orb-product}), here
\begin{equation} \label{eq:d4:fusion:start}
(2) \frac{1}{| {\mathbb Z}_2 |} \frac{1}{|D_4|^2}
\sum_{z \in {\mathbb Z}_2} \sum_{g_1h_1 = h_1g_1z}
\sum_{g_2h_2 = h_2g_2z}
\sum_{\gamma \in D_4}
 \epsilon_k(z)
\epsilon_{\ell}(z) 
\raisebox{-25pt}{
\begin{tikzpicture}
\draw (0,0) -- (0,0.5);  \draw (0,0) -- (0.5,-0.25);
\draw (0,0.5) -- (0.5,0.25);  \draw (0.5,0.25) -- (0.5,-0.25);
\draw (0,0.5) -- (1.5,0.5);  \draw (0.5,0.25) -- (2,0.25);
\draw (0.5,-0.25) -- (2,-0.25);
\draw (1.5,0.5) -- (2,0.25); \draw (2,0.25) -- (2,-0.25);
\node [below] at (1.25,-0.25) {$\gamma$};
\end{tikzpicture}
}
\end{equation}
where we have already identified the gerbe characteristic classes on
either end as a single $z \in H^2(T^2,{\mathbb Z}_2) = {\mathbb Z}_2$,
and where
\begin{equation}
g_1 \: = \: \gamma g_2 \gamma^{-1},
\: \: \:
h_1 \: = \: \gamma h_2 \gamma^{-1}.
\end{equation}

Counting $g_1, h_1, g_2, h_2, \gamma \in D_4$ such that, for any
fixed $z \in K \subset D_4$,
\begin{equation}
g_i h_i \: = \: h_i g_i z, \: \: \:
g_1 \: = \: \gamma g_2 \gamma^{-1}, \: \: \:
h_1 \: = \: \gamma h_2 \gamma^{-1}
\end{equation}
is a nontrivial combinatorial problem.
For example, in the $z=1$ sector, if we let $L$ and $R$ denote the
diagrams on either end of the box, then
\begin{equation}
L: \: {\scriptstyle 1} \square_z , \: \: \:
R: \: {\scriptstyle 1} \square_z
\end{equation}
are related by any $\gamma \in D_4$, since both $1, z$ commute with everything,
but only $\gamma \in \{b, bz, ab, abz\}$ can relate
\begin{equation}
L: \: {\scriptstyle a} \square_a, \: \: \:
R: \: {\scriptstyle az} \square_{az},
\end{equation}
and there are no $\gamma \in D_4$ at all that can relate
\begin{equation}
L: \: {\scriptstyle a} \square_a, \: \: \:
R: \: {\scriptstyle a} \square_{az}.
\end{equation}

In table~\ref{table:d4:zeq1-sector} 
we have compiled a list of prototypical examples 
of 
$L$, $R$ boundary configurations, ${\mathbb Z}_2 \times {\mathbb Z}_2$ orbifold
twisted sectors to which they project, and a count of the total numer of 
$(g_1, h_1, g_2, h_2, \gamma)$ of the form given, for the case $z=1$.
In table~\ref{table:d4:zneq1-sector} we have listed
analogous examples for the case $z \neq 1$.

Briefly, we find that in each case, for each ${\mathbb Z}_2 \times {\mathbb Z}_2$
sector, there are $(4)(8) = 32$ sets of $(g_1,h_1,g_2,h_2,\gamma)$ that
realize that sector.  The details of how those sectors are implemented
by $(g_1,h_1,g_2,h_2,\gamma)$ vary considerably between cases, as the
tables illustrate.

\begin{table}[h]
\begin{center}
\begin{tabular}{ccccc}
$L$ & $R$ & $\gamma$ & ${\mathbb Z}_2 \times {\mathbb Z}_2$ sector &
Total number \\ \hline
${\scriptstyle 1, z} \square_{1,z}$ & same & $D_4$ & ${\scriptstyle 1} \square_1$
& $(4)(8)$ \\
${\scriptstyle a, az} \square_{a,az}$ & same & $1, z, a, az$ &
${\scriptstyle \overline{a}} \square_{\overline{a}}$ & $(4)(4)$ \\
${\scriptstyle a} \square_a$ & ${\scriptstyle az} \square_a$ & none &
& $0$ \\
${\scriptstyle a,az} \square_{a,az}$ & ${\scriptstyle az,a} \square_{az,a}$
& $b, bz, ab, abz$ & ${\scriptstyle \overline{a}} \square_{\overline{a}}$
& $(4)(4)$\\
${\scriptstyle 1,z} \square_{a,az}$ & same & $1,z,a,az$ & 
${\scriptstyle 1} \square_{\overline{a}}$ & $(4)(4)$ \\
${\scriptstyle 1,z} \square_{a,az}$ &  ${\scriptstyle 1,z} \square_{az,a}$ &
$b,bz,ab,abz$ & ${\scriptstyle 1} \square_{\overline{a}}$ & $(4)(4)$ 
\end{tabular}
\caption{ \label{table:d4:zeq1-sector}
Listed here are some prototypical examples of twisted sectors $L$, $R$ 
on the boundary of the box, $\gamma \in D_4$ relating them,
${\mathbb Z}_2 \times {\mathbb Z}_2$ twisted sectors to which they project,
and the total number of $(g_1, h_1, g_2, h_2, \gamma) \in D_4^5$ of
this form that project to the same ${\mathbb Z}_2 \times {\mathbb Z}_2$
twisted sector, all for the case $z=1$.
}
\end{center}
\end{table}

\begin{table}[h]
\begin{center}
\begin{tabular}{ccccc}
$L$ & $R$ & $\gamma$ & ${\mathbb Z}_2 \times {\mathbb Z}_2$ sector &
Total number \\ \hline
${\scriptstyle a,az} \square_{b,bz}$ & same & $1,z$ & 
${\scriptstyle \overline{a}} \square_{\overline{b}}$ & $(4)(2)$
\\
${\scriptstyle a,az} \square_{b,bz}$ & ${\scriptstyle az,a} \square_{bz,b}$
& none & & $0$ 
\\
${\scriptstyle a,az} \square_{b,bz}$ & ${\scriptstyle a,az} \square_{bz,b}$
& $a,az$ & ${\scriptstyle \overline{a}} \square_{\overline{b}}$ & $(4)(2)$
\\
${\scriptstyle a,az} \square_{b,bz}$ & ${\scriptstyle az,a} \square_{b,bz}$
& $b,bz,ab,abz$ & ${\scriptstyle \overline{a}} \square_{\overline{b}}$ &
$(4)(4)$
\end{tabular}
\caption{ \label{table:d4:zneq1-sector}
Listed here are some prototypical examples of twisted sectors $L$, $R$ 
on the boundary of the box, $\gamma \in D_4$ relating them,
${\mathbb Z}_2 \times {\mathbb Z}_2$ twisted sectors to which they project,
and the total number of $(g_1, h_1, g_2, h_2, \gamma) \in D_4^5$ of
this form that project to the same ${\mathbb Z}_2 \times {\mathbb Z}_2$
twisted sector, all for the case $z \neq 1$.
}
\end{center}
\end{table}

Now, we can simplify the partition function of the
fusion product~(\ref{eq:d4:fusion:start},
following the same analysis as in \cite{Sharpe:2019ddn}.
First,
\begin{equation}
\epsilon_k(z) \, \epsilon_{\ell}(z) \: = \: \epsilon_{k+\ell}(z).
\end{equation}
From the same reasoning as in \cite{Sharpe:2019ddn},
when $k + \ell = 0 \mod 2$, this will simply add together the
$z=1$ and $z=-1$ sectors, but when $k + \ell = 1 \mod 2$, this will
have the effect of multiplying some of the ${\mathbb Z}_2 \times
{\mathbb Z}_2$ sectors by discrete torsion (as precisely the
$z=-1$ sectors are the ones that are weighted by signs under
${\mathbb Z}_2 \times {\mathbb Z}_2$ discrete torsion).
Putting this together, we see that
\begin{eqnarray}
\lefteqn{
(2) \frac{1}{| {\mathbb Z}_2 |} \frac{1}{|D_4|^2}
\sum_{z \in {\mathbb Z}_2} \sum_{g_1h_1 = h_1g_1z}
\sum_{g_2h_2 = h_2g_2z}
\sum_{\gamma \in D_4}
 \epsilon_k(z)
\epsilon_{\ell}(z) 
\raisebox{-25pt}{
\begin{tikzpicture}
\draw (0,0) -- (0,0.5);  \draw (0,0) -- (0.5,-0.25);
\draw (0,0.5) -- (0.5,0.25);  \draw (0.5,0.25) -- (0.5,-0.25);
\draw (0,0.5) -- (1.5,0.5);  \draw (0.5,0.25) -- (2,0.25);
\draw (0.5,-0.25) -- (2,-0.25);
\draw (1.5,0.5) -- (2,0.25); \draw (2,0.25) -- (2,-0.25);
\node [below] at (1.25,-0.25) {$\gamma$};
\end{tikzpicture}
}
} \nonumber \\
& = & \left\{ \begin{array}{cl}
2 \, Z_{T^2}\left( [X/{\mathbb Z}_2 \times {\mathbb Z}_2] \right)
\: = \: 2 \, Z\left(S_0(\Sigma)\right)
& k + \ell \equiv 0 \mod 2,
\\
2 \,
Z_{T^2}\left( [X/{\mathbb Z}_2 \times {\mathbb Z}_2]_{d.t.} \right)
\: = \: 2 \, Z\left( S_1(\Sigma) \right)
& k + \ell \equiv 1 \mod 2,
\end{array} \right.
\end{eqnarray}
where the d.t. subscript indicates that the ${\mathbb Z}_2 \times
{\mathbb Z}_2$ orbifold is computed with discrete torsion.

The factor of 2 should be interpreted to mean that the fusion product
$S_i(\Sigma) \times S_j(\Sigma)$,
is two copies, either
\begin{equation}
[X / {\mathbb Z}_2 \times {\mathbb Z}_2 ] |_{\Sigma} \, \coprod \,
[X / {\mathbb Z}_2 \times {\mathbb Z}_2 ] |_{\Sigma}
\end{equation}
or
\begin{equation}
[X / {\mathbb Z}_2 \times {\mathbb Z}_2 ]_{\rm d.t.} |_{\Sigma} \, \coprod \,
[X / {\mathbb Z}_2 \times {\mathbb Z}_2 ]_{\rm d.t.} |_{\Sigma}
\end{equation}
along $\Sigma$, or in other words,
\begin{eqnarray}
S_0(\Sigma) \times S_0(\Sigma) & = & 2 \, S_0(\Sigma),
\\
S_0(\Sigma) \times S_1(\Sigma) & = & 2 \, S_1(\Sigma),
\\
S_1(\Sigma) \times S_1(\Sigma) & = & 2 \, S_0(\Sigma).
\end{eqnarray}

These defects arose from gauging a $B {\mathbb Z}_2$ in an orbifold
with a trivially-acting ${\mathbb Z}_2$, closely related to
the example discussed in \cite[section 6.3]{Roumpedakis:2022aik}
and our section~\ref{sect:z2}.
There, as was previously observed in our section~\ref{sect:z2},
the analogue of the
defect that is labelled ``$S_e$'' in \cite[section 6.3]{Roumpedakis:2022aik}
is here two distinct, albeit isomorphic, defects.  
In this example, the distinction between those two defects $S_0(\Sigma)$,
$S_1(\Sigma)$ is much more clear.

The fusion rule obtained in \cite[section 6.3]{Roumpedakis:2022aik} 
for $S_e$ was 
simply
\begin{equation}
S_e \times S_e \: = \: S_e + S_e.
\end{equation}
The fusion rules we have derived above for $S_{0,1}(\Sigma)$ are therefore
of the expected form, as they refine the fusion rule for $S_e$
in section~\ref{sect:z2}, analogous to fusion rules for
condensation defects in \cite[section 6.3]{Roumpedakis:2022aik}.
(The reader may find it useful to recall from section~\ref{sect:mult}
that $S_e + S_e$ is equivalent to coupling $S_e$ to a particular
topological field theory.)

\subsubsection{Example:  $[X/{\mathbb H}]$}
\label{sect:h}

Now, consider the three-dimensional orbifold $[X/{\mathbb H}]$,
where ${\mathbb H}$ is the eight-element group of unit quaternions,
and $\langle i \rangle \cong {\mathbb Z}_4 \subset {\mathbb H}$
acts trivially on $X$.

This three-dimensional theory has a one-form symmetry, and its
restriction to a two-dimensional submanifold $\Sigma$ of spacetime
decomposes, as \cite[section 5.4]{Hellerman:2006zs}
\begin{equation}
[X/{\mathbb H}] |_{\Sigma} \: = \:
X |_{\Sigma} \, \coprod \,
[X/{\mathbb Z}_2] |_{\Sigma} \, \coprod \,
[X/{\mathbb Z}_2] |_{\Sigma}.
\end{equation}
Because ${\mathbb Z}_4$ is not in the center, part of that one-form symmetry
is realized non-invertibly, as discussed in \cite{Sharpe:2021srf}.
That trivially-acting ${\mathbb Z}_4$ contains the ${\mathbb Z}_2$ center
of ${\mathbb H}$, and the $B {\mathbb Z}_2$ is realized linearly.

Consider gauging that $B {\mathbb Z}_2$ symmetry along $\Sigma$.
Applying a slight variant\footnote{
Reference \cite[section 6.3]{Sharpe:2019ddn} formally considered gauging the
$B {\mathbb Z}_4$, not just the ${\mathbb Z}_2$.
The analysis for $B {\mathbb Z}_2$ is nearly identical, the only real
change is to replace the $1/| {\mathbb Z}_4|$ factor with
$1/|{\mathbb Z}_2|$, as only $z = \pm 1$ contribute to the sum over gerbes.
The results of the $B {\mathbb Z}_2$ gauging are as indicated above.
} of the analysis in
\cite[section 6.3]{Sharpe:2019ddn}, reviewed in section~\ref{sect:2dorb:rev},
by gauging a $B {\mathbb Z}_2$, one gets (depending upon the one-form
theta angle) either 
\begin{equation}
[X/{\mathbb Z}_2] |_{\Sigma} \, \coprod \,
[X/{\mathbb Z}_2] |_{\Sigma}
\end{equation}
(for $\epsilon(z=-1) = +1$) or
\begin{equation}
X |_{\Sigma}
\end{equation}
(for $\epsilon(z=-1) = -1$).
Denote the two resulting defects by $S_{0,1}(\Sigma)$:
\begin{equation}
S_0(\Sigma) \: = \: [X/{\mathbb Z}_2] |_{\Sigma} \, \coprod \,
[X/{\mathbb Z}_2] |_{\Sigma},
\: \: \:
S_1(\Sigma) \: = \: X |_{\Sigma}.
\end{equation}
In this case, $S_0(\Sigma)$ is reducible, though we will not utilize that
fact.

Next, let us compute the fusion product of these defects.
Take $\Sigma = T^2$, then from~(\ref{eq:orb-product}), 
we have that the partition function
of the fusion is of the form
\begin{equation}
(2) \frac{ 1 }{ | {\mathbb Z}_2 | } \frac{1}{| {\mathbb H} |^2}
\sum_{z \in {\mathbb Z}_2} \sum_{g_1 h_1 = h_1 g_1 z} \sum_{g_2 h_2 = h_2 g_2 z}
\sum_{\gamma \in {\mathbb H}}
\epsilon_{\ell_1}(z) \epsilon_{\ell_2}(z)
\raisebox{-25pt}{
\begin{tikzpicture}
\draw (0,0) -- (0,0.5);  \draw (0,0) -- (0.5,-0.25);
\draw (0,0.5) -- (0.5,0.25);  \draw (0.5,0.25) -- (0.5,-0.25);
\draw (0,0.5) -- (1.5,0.5);  \draw (0.5,0.25) -- (2,0.25);
\draw (0.5,-0.25) -- (2,-0.25);
\draw (1.5,0.5) -- (2,0.25); \draw (2,0.25) -- (2,-0.25);
\node [below] at (1.25,-0.25) {$\gamma$};
\end{tikzpicture}
}
\end{equation}
where
\begin{equation}
g_1 \: = \: \gamma g_2 \gamma^{-1}, \: \: \:
h_1 \: = \: \gamma h_2 \gamma^{-1}.
\end{equation}

As before, counting collections $(g_1,h_1,g_2,h_2,\gamma)$ satisfying
the conditions above for any fixed $z$ is an exercise in combinatorics.
In tables~\ref{table:h:sectors:zeq1}, \ref{table:h:sectors:zneq1} 
we have summarized results for some pertinent cases, and a summary
of the sector counting is given in table~\ref{table:h:summ}.
In the tables, $\xi$ denotes the generator of the effective
${\mathbb Z}_2 = {\mathbb H}/\langle i \rangle$ orbifold.

\begin{table}[h]
\begin{center}
\begin{tabular}{ccc|cc}
 $(g_1, h_1)$ & $(g_2,h_2)$ & $\gamma$ & ${\mathbb Z}_2$ sector & Number \\
\hline
 $(\pm 1, \pm 1)$ & same & all & $(1,1)$ & $(4)(8)$ \\
 $(\pm 1, \pm i)$ & same & $\pm 1, \pm i$ & $(1,1)$ & $(4)(4)$ \\
 $(\pm 1, \pm i)$ & $(\pm 1, \mp i)$ & $\pm j, \pm k$ & $(1,1)$ & $(4)(4)$ \\
 $(\pm i, \pm 1)$ & same & $\pm 1, \pm i$ & $(1,1)$ & $(4)(4)$ \\
 $(\pm i, \pm 1)$ & $(\mp i, \pm 1)$ & $\pm j, \pm k$ & $(1,1)$ & $(4)(4)$ \\
 $(\pm i, \pm i)$ & same & $\pm 1, \pm i$ & $(1,1)$ & $(4)(4)$ \\
 $(\pm i, \pm i)$ & $(\mp i, \mp i)$ & $\pm j, \pm k$ & $(1,1)$ & $(4)(4)$ \\
\hline
 $(\pm j, \pm j)$ & same & $\pm 1, \pm j$ & $(\xi, \xi)$ & $(4)(4)$ \\
 $(\pm j, \pm j)$ & $(\mp j, \mp j)$ & $\pm i, \pm k$ & $(\xi,\xi)$ & $(4)(4)$
\\
 $(\pm k, \pm k)$ & same & $\pm 1, \pm k$ & $(\xi,\xi)$ & $(4)(4)$ \\
 $(\pm k, \pm k)$ & $(\mp k, \mp k)$ & $\pm i, \pm j$ & $(\xi,\xi)$ & $(4)(4)$
\end{tabular}
\caption{\label{table:h:sectors:zeq1}
Some prototypical examples of twisted sectors on the boundary of the box
defining a fusion product of defects on $\Sigma = T^2$ in $[X/{\mathbb H}]$,
for $z = +1$.
}
\end{center}
\end{table}

\begin{table}[h]
\begin{center}
\begin{tabular}{ccc|cc}
 $(g_1, h_1)$ & $(g_2,h_2)$ & $\gamma$ & ${\mathbb Z}_2$ sector & Number \\
\hline
 $(\pm 1, \pm j)$ & same & $\pm 1, \pm j$ & $(1,\xi)$ & $(4)(4)$ \\
 $(\pm 1, \pm j)$ & $(\pm 1, \mp j)$ & $\pm i, \pm k$ & $(1,\xi)$ & $(4)(4)$ \\
 $(\pm 1, \pm k)$ & same & $\pm 1, \pm k$ & $(1,\xi)$ & $(4)(4)$ \\
 $(\pm 1, \pm k)$ & $(\pm 1, \mp k)$ & $\pm i, \pm j$ & $(1,\xi)$ & $(4)(4)$\\
\hline
 $(\pm i, \pm j)$ & same & $\pm 1$ & $(1,\xi)$ & $(4)(2)$ \\
 $(\pm i, \pm j)$ & $(\pm i, \mp j)$ & $\pm i$ & $(1,\xi)$ & $(4)(2)$ \\
 $(\pm i, \pm j)$ & $(\mp i, \pm j)$ & $\pm j$ & $(1,\xi)$ & $(4)(2)$ \\
 $(\pm i, \pm j)$ & $(\mp i, \mp j)$ & $\pm k$ & $(1,\xi)$ & $(4)(2)$ \\
\hline
 $(\pm i, \pm k)$ & same & $\pm 1$ & $(1,\xi)$ & $(4)(2)$ \\
 $(\pm i, \pm k)$ & $(\pm i, \mp k)$ & $\pm i$ & $(1,\xi)$ & $(4)(2)$ \\
 $(\pm i, \pm k)$ & $(\mp i, \pm k)$ & $\pm k$ & $(1,\xi)$ & $(4)(2)$ \\
 $(\pm i, \pm k)$ & $(\mp i, \mp k)$ & $\pm j$ & $(1,\xi)$ & $(4)(2)$ \\
\hline
 $(\pm j, \pm k)$ & same & $\pm 1$ & $(\xi,\xi)$ & $(4)(2)$ \\
 $(\pm j, \pm k)$ & $(\pm j, \mp k)$ & $\pm j$ & $(\xi,\xi)$ & $(4)(2)$ \\
 $(\pm j, \pm k)$ & $(\mp j, \pm k)$ & $\pm k$ & $(\xi,\xi)$ & $(4)(2)$ \\
 $(\pm j, \pm k)$ & $(\mp j, \mp k)$ & $\pm i$ & $(\xi,\xi)$ & $(4)(2)$
\end{tabular}
\caption{\label{table:h:sectors:zneq1}
Some prototypical examples of twisted sectors on the boundary of the box
defining a fusion product of defects on $\Sigma = T^2$ in $[X/{\mathbb H}]$,
for $z = -1$.
}
\end{center}
\end{table}

\begin{table}[h]
\begin{center}
\begin{tabular}{ccc}
${\mathbb Z}_2$ sector & Num. appearances in $z=+1$ & Num. appearances in $z=-1$
\\ \hline
$(1,1)$ & $128$ & $0$ \\
$(1,\xi)$ & $64$ & $64$ \\
$(\xi,1)$ & $64$ & $64$ \\
$(\xi,\xi)$ & $64$ & $64$ 
\end{tabular}
\caption{\label{table:h:summ}
A summary of the counting of twisted sectors appearing in the fusion product
of defects in $[X/{\mathbb H}]$
and their relation to ${\mathbb Z}_2$ orbifold sectors.
}
\end{center}
\end{table}

Now, we can assemble these pieces.  From table~\ref{table:h:summ},
we see that the partition function of the fusion product is given by
\begin{eqnarray}
\lefteqn{
(2) \frac{ 1 }{ | {\mathbb Z}_2 | } \frac{1}{| {\mathbb H} |^2}
\sum_{z \in {\mathbb Z}_2} \sum_{g_1 h_1 = h_1 g_1 z} \sum_{g_2 h_2 = h_2 g_2 z}
\sum_{\gamma \in {\mathbb H}}
\epsilon_{\ell_1}(z) \epsilon_{\ell_2}(z)
\raisebox{-25pt}{
\begin{tikzpicture}
\draw (0,0) -- (0,0.5);  \draw (0,0) -- (0.5,-0.25);
\draw (0,0.5) -- (0.5,0.25);  \draw (0.5,0.25) -- (0.5,-0.25);
\draw (0,0.5) -- (1.5,0.5);  \draw (0.5,0.25) -- (2,0.25);
\draw (0.5,-0.25) -- (2,-0.25);
\draw (1.5,0.5) -- (2,0.25); \draw (2,0.25) -- (2,-0.25);
\node [below] at (1.25,-0.25) {$\gamma$};
\end{tikzpicture}
}
} \nonumber \\
& = &
\left\{ \begin{array}{cl}
4 \, Z_{T^2}\left( [X/{\mathbb Z}_2] \right) = 2 \, Z\left( S_0(\Sigma) \right)
& \ell_1 + \ell_2 = 0 \mod 2,
\\
2 \, Z_{T^2}(X) \: = \: 2 \, Z\left( S_1(\Sigma) \right)
& \ell_1 + \ell_2 = 1 \mod 2.
\end{array} \right.
\end{eqnarray}
Put more simply, this implies the fusion rules
\begin{eqnarray}
S_0(\Sigma) \times S_0(\Sigma) & = & 2 \, S_0(\Sigma),
\\
S_0(\Sigma) \times S_1(\Sigma) & = & 2 \, S_1(\Sigma),
\\
S_1(\Sigma) \times S_1(\Sigma) & = & 2 \, S_0(\Sigma),
\end{eqnarray}
of the form expected from results in sections~\ref{sect:z2}, \ref{sect:d4},
and similar to results for analogous condensation defects in
\cite[section 6.3]{Roumpedakis:2022aik}.

\subsubsection{Example:  ${\mathbb Z}_p$ gauge theory}
\label{sect:zp}

Let us now consider the case that the three-dimensional orbifold is
$[X/{\mathbb Z}_p]$ with all of the ${\mathbb Z}_p$ acting trivially,
as in \cite[section 6.4]{Roumpedakis:2022aik}.
This theory has a $B {\mathbb Z}_p$ symmetry, and its restriction to
a two-dimensional $\Sigma$ therefore decomposes, in this case to
$p$ identical copies of a sigma model on $X$.  If we gauge $B {\mathbb Z}_n$
along $\Sigma$, for $n$ a divisor of $p$, then depending upon the choice of
one-form theta angle, we will recover subsets of that collection of
sigma models, consisting of sums of $p/n$ copies of sigma models on $X$.
We will denote those defects $S_{p,n,k}(\Sigma)$,
where $n$ divides $p$ (corresponding to the gauged $B {\mathbb Z}_n$)
and $k \in \{0, \cdots, n-1\}$, indexing the various copies.

As before, suppose that $\Sigma = T^2$, so that in the large tension limit
the partition function for
$S_{p,n,k}(\Sigma)$ is \cite[equ'n (6.9)]{Sharpe:2019ddn}
\begin{equation}
Z\left( S_{p,n,k}(\Sigma) \right) 
\: = \: \frac{1}{ | {\mathbb Z}_n | } \frac{1}{| {\mathbb Z}_p|}
\sum_{z \in {\mathbb Z}_n} \sum_{gh=hgz} \epsilon_k(z)
{\scriptstyle g} \tsquare{z}_h.
\end{equation}
Here, as ${\mathbb Z}_p$ is abelian, there are no contributions when
$z \neq 1$, and as $\epsilon_k(+1) = +1$ for all $k$, this reduces to
\begin{equation}
Z\left( S_{p,n,k}(\Sigma) \right) \: = \:
\frac{1}{| {\mathbb Z}_n | } \frac{1}{| {\mathbb Z}_p |}
\sum_{gh = hg} {\scriptstyle g} \square_h
\: = \: (p/n) Z(\Sigma, X)
\end{equation}
for all values of $k$, corresponding to $p/n$ copies of a sigma model on $X$.
In particular, this suggests that
\begin{equation}
S_{p,n,i}(\Sigma) \: \cong \: S_{p,n,j}(\Sigma) 
\: \cong \: \oplus^{p/n} X |_{\Sigma}
\end{equation}
for all $i, j \in \{0, \cdots, n-1 \}$.
Since the result is independent of the last index, we will sometimes write
each of these defects as $S_{p,n}(\Sigma)$, omitting the last index.

Now, let us compute the fusion product, when one defect has a gauged
$B {\mathbb Z}_n$, and the other a gauged $B {\mathbb Z}_{n'}$.
From~(\ref{eq:orb-product}), the partition function of the fusion
product is
\begin{equation}
{\rm gcd}(n,n')
\frac{1}{| {\mathbb Z}_{ {\rm lcm}(n,n') } | }
\frac{1}{| {\mathbb Z}_p|^2}
\sum_{z \in {\mathbb Z}_{ {\rm lcm}(n,n')} }
\sum_{g_1 h_1 = h_1 g_1 z}
\sum_{g_2 h_2 = h_2 g_2 z} 
\sum_{\gamma \in {\mathbb Z}_p}
\epsilon_{\ell_1}(z) \epsilon_{\ell_2}(z)
\raisebox{-25pt}{
\begin{tikzpicture}
\draw (0,0) -- (0,0.5);  \draw (0,0) -- (0.5,-0.25);
\draw (0,0.5) -- (0.5,0.25);  \draw (0.5,0.25) -- (0.5,-0.25);
\draw (0,0.5) -- (1.5,0.5);  \draw (0.5,0.25) -- (2,0.25);
\draw (0.5,-0.25) -- (2,-0.25);
\draw (1.5,0.5) -- (2,0.25); \draw (2,0.25) -- (2,-0.25);
\node [below] at (1.25,-0.25) {$\gamma$};
\end{tikzpicture}
}.
\end{equation}
In this case, since $\Gamma$ is abelian, the only solutions of
\begin{equation}
g_i h_i \: = \: h_i g_i z
\end{equation}
require $z=1$, and as $\epsilon_{\ell}(z=1) = +1$, the $\epsilon$
factors drop out.
Furthermore, since $\Gamma$ is abelian,
\begin{equation}
g_1 \: = \: \gamma g_2 \gamma^{-1} \: = \: g_2,
\: \: \:
h_1 \: = \: \gamma h_2 \gamma^{-1} : = \: h_2,
\end{equation}
so we see that $g_2$, $h_2$ are uniquely determined by $g_1$, $h_1$,
and the sum over $\gamma$ merely contributes an overall factor
of $| \Gamma| = | {\mathbb Z}_p|$.  Putting this together, we find
\begin{eqnarray}
\lefteqn{
{\rm gcd}(n,n')
\frac{1}{| {\mathbb Z}_{ {\rm lcm}(n,n') } | }
\frac{1}{| {\mathbb Z}_p|^2}
\sum_{z \in {\mathbb Z}_{ {\rm lcm}(n,n')} }
\sum_{g_1 h_1 = h_1 g_1 z}
\sum_{g_2 h_2 = h_2 g_2 z} 
\sum_{\gamma \in {\mathbb Z}_p}
\epsilon_{\ell_1}(z) \epsilon_{\ell_2}(z)
\raisebox{-25pt}{
\begin{tikzpicture}
\draw (0,0) -- (0,0.5);  \draw (0,0) -- (0.5,-0.25);
\draw (0,0.5) -- (0.5,0.25);  \draw (0.5,0.25) -- (0.5,-0.25);
\draw (0,0.5) -- (1.5,0.5);  \draw (0.5,0.25) -- (2,0.25);
\draw (0.5,-0.25) -- (2,-0.25);
\draw (1.5,0.5) -- (2,0.25); \draw (2,0.25) -- (2,-0.25);
\node [below] at (1.25,-0.25) {$\gamma$};
\end{tikzpicture}
}
} \nonumber \\
& \hspace*{0.75in} = &
{\rm gcd}(n,n')
\frac{1}{| {\mathbb Z}_{ {\rm lcm}(n,n') } | }
\frac{1}{| {\mathbb Z}_p|}
\sum_{g_1, h_1} {\scriptstyle g_1} \square_{h_1} \, ,
\\
& \hspace*{0.75in} = &
{\rm gcd}(n,n')
\frac{ | {\mathbb Z}_p |}{| {\mathbb Z}_{ {\rm lcm}(n,n') } | }
{\scriptstyle 1} \square_1
\: = \:
{\rm gcd}(n,n')
\frac{ | {\mathbb Z}_p |}{| {\mathbb Z}_{ {\rm lcm}(n,n') } | }
Z(\Sigma, X).
\end{eqnarray}

The computation above suggests that
the fusion product is
\begin{equation}  \label{eq:zp:fusion}
S_{p,n} \times S_{p,n'} \: = \:
\oplus^{ (p) {\rm gcd}(n,n') / {\rm lcm}(n,n') } X |_{\Sigma},
\end{equation}
which can be rewritten in terms of the $S_{p,m}$ for various $m$
depending upon $p$, $n$, $n'$.
The reader should note that since $n$ and $n'$ both divide $p$,
the ratio
\begin{equation}
p \, \frac{ {\rm gcd}(n,n') }{ {\rm lcm}(n,n') }
\end{equation}
is a positive integer.

Now, let us compare to the results for analogous
condensation defects in \cite[section 6.4]{Roumpedakis:2022aik},
which also considered ${\mathbb Z}_p$ gauge theories in three
dimensions, for $p$ prime.
The defects above correspond, in the language of
\cite[section 6.4]{Roumpedakis:2022aik}, to gauging a single cyclic factor,
hence their $m = \infty$.  If we take $p$ to be prime and $n = n' = p$,
so that, for example,
\begin{equation}
S_{p,n}(\Sigma) \: = S_{p,n'}(\Sigma) \: = \: X |_{\Sigma},
\end{equation}
then our result~(\ref{eq:zp:fusion}) reduces to
\begin{equation}
S_{ {\mathbb Z}_p^{(\infty)} } \times
S_{ {\mathbb Z}_p^{(\infty)} } \: = \: p X |_{\Sigma} \: = \:
p S_{ {\mathbb Z}_p^{(\infty)} },
\end{equation}
in the notation of \cite[section 6.4]{Roumpedakis:2022aik},
which matches the pertinent piece of \cite[equ'n (6.70)]{Roumpedakis:2022aik},
after taking into account the relation between topological field theory
factors and multiplicities explained in section~\ref{sect:mult}.

To be clear, our analysis is somewhat orthogonal to that of
\cite[section 6.4]{Roumpedakis:2022aik}, both because the
defects here are not, so far as we are aware, condensation defects,
and also because \cite[section 6.4]{Roumpedakis:2022aik}
considers more general
gaugings (and hence more general defects) for the case of $p$ prime
than we have
considered here.

\subsection{Three-dimensional defects in orbifolds by 2-groups}

In this section we discuss three-dimensional defects,
in higher-dimensional orbifolds by 2-groups.  We begin with an overview
of three-dimensional orbifolds by 2-groups, their decomposition,
as well as previously unpublished results on gauging global two-form
symmetries in such theories, then we turn to a study of defects
specifically.

\subsubsection{Orbifolds  by 2-groups in three dimensions} 
\label{sect:3dorb}

Decomposition in three-dimensional orbifolds by 2-groups
was discussed in
\cite{Pantev:2022kpl}.  Specifically, that work discussed
three-dimensional orbifolds $[X/\tilde{\Gamma}]$ where
$\tilde{\Gamma}$ is a two-group extension of an ordinary finite group $G$
by a trivially-acting one-form group $BK$:
\begin{equation}  \label{eq:2gp:ext}
1 \: \longrightarrow \: BK \: \longrightarrow \: \tilde{\Gamma} \: 
\longrightarrow \: G \: \longrightarrow \: 1.
\end{equation}
Since the gauged $BK$ acts trivially, the theory has a global
2-form symmetry, and hence decomposes.
Specifically, it was argued that
\begin{equation}  \label{eq:3d-decomp}
{\rm QFT}\left( [X/\tilde{\Gamma}]\right) \: = \:
\coprod_{\rho \in \hat{K}} {\rm QFT}\left( [X/G]_{\rho(\omega)} \right),
\end{equation}
where $\omega \in H^3(G,K)$ corresponds to the extension class
of~(\ref{eq:2gp:ext}), $\rho(\omega) \in H^3(G,K)$ is the
composition of $\rho \in \hat{K}$ with the extension class $\omega$,
and the orbifold $[X/G]$ is twisted by a three-dimensional
analogue of discrete torsion, as in e.g.~\cite{Dijkgraaf:1989pz,Sharpe:2000qt}.

If the three-manifold $Y = T^3$, then the partition function of
$[X/\tilde{\Gamma}]$
is \cite[equ'n (4.18)]{Pantev:2022kpl}
\begin{eqnarray}  \label{eq:3d:2gp:genl-part-fn}
Z_{T^3}\left( [X/\tilde{\Gamma} ] \right)
& = &
\frac{ | H^0(T^3, K) | }{ | H^1(T^3, K) | } \frac{1}{|H^0(T^3,G)|}
 \sum_{z_1, z_2, z_3 \in K}
{\sum_{g_1, g_2, g_3 \in G}}^{\!\!\!\!\!\prime} \: Z(g_1, g_2, g_3),
\nonumber\\
& = &
\frac{1}{|K|^2 |G|} \sum_{z_1, z_2, z_3 \in K}
{\sum_{g_1, g_2, g_3 \in G}}^{\!\!\!\!\!\prime}\: Z(g_1, g_2, g_3),
\end{eqnarray}
where the prime ($\prime$) indicates that the sum over elements of $G$
is restricted to those such that
\begin{equation}
\epsilon_Y(g_1,g_2,g_3) \: = \: 1,
\end{equation}
reflecting the fact that not all $G$ bundles appear as images of
$\tilde{\Gamma}$ orbifolds, as discussed in \cite{Pantev:2022kpl}.
For $Y = T^3$,
\begin{equation}
\epsilon_{Y=T^3}(g_1, g_2, g_3) \: = \: 
\frac{ \omega(g_1, g_2, g_3) }{ \omega(g_1, g_3, g_2) }
\frac{ \omega(g_3, g_1, g_2) }{ \omega(g_3, g_2, g_1) }
\frac{ \omega(g_2, g_3, g_1) }{ \omega(g_2, g_1, g_3) },
\end{equation}
where $\omega \in H^3(G,K)$ is the class of the extension~(\ref{eq:2gp:ext}).
Projecting to $G$ bundles of that form is equivalent to working with
a sum over universes, and it was argued in \cite{Pantev:2022kpl} that
\begin{equation}
Z_{T^3}\left( [X/\tilde{\Gamma} ] \right)
\: = \:
\sum_{\rho \in \hat{K}}  Z_{T^3}\left( [X/G]_{\epsilon_{\rho})} \right),
\end{equation}
reflecting the decomposition~(\ref{eq:3d-decomp}).

Just as in the case of decomposing two-dimensional theories
discussed in \cite{Sharpe:2019ddn}, where one gauges the global
one-form symmetry to recover individual universes,
in principle in a decomposing three-dimensional theory one should be able
to gauge the global two-form symmetry to recover individual universes.
In this section we will review results on decomposition in
such three-dimensional orbifolds, and also suggest a concrete
mechanism to gauge the global two-form symmetry so as to recover
individual universes in the decomposition, results we shall be
utilizing later in this paper.

Let us describe the partition function of such a gauging explicitly.
In general terms, the partition function of $[ [X/\tilde{\Gamma}] / B^2 K]$
on a three-manifold $Y$ should have the form
\begin{eqnarray} 
\lefteqn{
Z_Y\left( \left[ [X/\tilde{\Gamma}] / B^2 K\right] \right)
} \nonumber \\
& = & \frac{1}{|K|} \sum_{\gamma \in H^3(Y,K)} \epsilon_R(\gamma)
\left\{ \mbox{sum over $\gamma$-twisted $\tilde{\Gamma}$ bundles and maps into
$X$} \right\},
\label{eq:gauged-b2k-2gp-orb}
\end{eqnarray}
where $\epsilon_R(\gamma)$ is the gauged two-form theta angle
associated with $R \in \hat{K}$ that determines which universe will
be selected.

It is natural to conjecture that a $\gamma$-twisted $\tilde{\Gamma}$ bundle on $Y$
defines a $G$ bundle on $Y$ obeying the constraint
\begin{equation}
\epsilon_Y(P) \: = \: \gamma,
\end{equation}
and we will see that this correctly selects out universes when
gauging $B^2K$ in a three-dimensional theory.

Utilizing the conjecture above,
our general formula~(\ref{eq:gauged-b2k-2gp-orb})
for the
partition function of a $B^2K$ gauged 2-group orbifold $[X/\tilde{\Gamma}]$
on the three-manifold $Y = T^3$ takes the form
\begin{eqnarray} 
\lefteqn{
Z_Y\left( \left[ [X/\tilde{\Gamma}] / B^2 K\right] \right)
} \nonumber \\
& = & \frac{1}{|K|} \sum_{\gamma \in H^3(Y,K)} \epsilon_R(\gamma)
\left[
\frac{1}{|K|^2 |G|} \sum_{z_{1-3} \in K} \sum_{g_{1-3} \in G, \epsilon_Y = \gamma}
Z(g_1, g_2, g_3) \right],
\label{eq:gauged-b2k-2gp-orb-t3}
\end{eqnarray}
where the sum in $G$ is
over commuting triples of elements of $G$ (such that $\epsilon_Y(g_1,g_2,g_3)
= \gamma$).

Let us check this in some simple examples.
First, from \cite[section 4.3]{Pantev:2022kpl},
consider the case $G={\mathbb Z}_2 = K$, so that
\begin{equation}
1 \: \longrightarrow \: B {\mathbb Z}_2 \: \longrightarrow \: \tilde{\Gamma}
\: \longrightarrow \: {\mathbb Z}_2 \: \longrightarrow \: 1.
\end{equation}
As noted in \cite[section 4.3]{Pantev:2022kpl}, there is a nontrivial
extension of this form, which we take.
For that extension, it was argued in \cite[section 4.3]{Pantev:2022kpl}
that the theory decomposes as
\begin{equation}
{\rm QFT}\left( [X/\tilde{\Gamma}] \right) \: = \:
\coprod_{\rho \in \hat{K}} {\rm QFT}\left( [X/G] \right),
\end{equation}
all universes with trivial discrete torsion.

Now, we consider partition functions on $Y = T^3$.
As noted in \cite[section 4.3]{Pantev:2022kpl}, 
for all $G$ bundles on $Y = T^3$ (meaning, commuting triples $(g_1, g_2, g_3)
\in G^3$),
\begin{equation}
\epsilon_Y(g_1,g_2,g_3) \: = \: 1,
\end{equation}
regardless of the extension class $\omega$.  As a result, there is
no constraint on $G$ bundles appearing in the partition function of
$[X/\tilde{\Gamma}]$ on $T^3$, and the orbifolds $[X/G]$ appearing in the
decomposition do not have any discrete torsion.
Explicitly, the partition
function of $[X/\tilde{\Gamma}]$ on $Y = T^3$ was given by
\begin{eqnarray}
Z_{T^3}\left( [X/\tilde{\Gamma}] \right) & = &
\frac{1}{|K|^2 |G|} \sum_{z_1, z_2, z_3 \in K} \sum_{g_1, g_2, g_3 \in G}
Z(g_1, g_2, g_3),
\\
& = & \frac{|K|}{|G|} \sum_{g_1, g_2, g_3} Z(g_1, g_2, g_3),
\\
& = & Z_{T^3}\left( \coprod_{\rho \in \hat{K}}  [X/G] \right),
\end{eqnarray}
consistent with the decomposition of $[X/\tilde{\Gamma}]$

Next, we consider gauging the global $B^2 K$ symmetry of this theory.
From the prescription we outlined above,
the partition function on $Y = T^3$ is given by
\begin{eqnarray}
\lefteqn{
Z_{T^3}\left( \left[ [X/\tilde{\Gamma}] / B^2 K \right] \right)
} \nonumber \\
& = &
\frac{1}{|K|} \sum_{\gamma \in H^3(T^3,K) = K} \epsilon_R(\gamma)
\left( \frac{1}{|K|^2 |G|} \sum_{z_1,z_2,z_3 \in K}
\sum_{g_1, g_2, g_3 \in G, \epsilon_Y = \gamma} Z(g_1, g_2, g_3)
\right).
\end{eqnarray}
Since $\epsilon_Y(g_1,g_2,g_3) = \gamma$ only has solutions in the
case $\gamma = 1$, we see that the only contributions to the path
integral arise from $\gamma = 1$, hence
\begin{eqnarray}
Z_{T^3}\left( \left[ [X/\tilde{\Gamma}] / B^2 K \right] \right)
& = &
\frac{1}{|K|} 
\left( \frac{1}{|K|^2 |G|} \sum_{z_1,z_2,z_3 \in K}
\sum_{g_1, g_2, g_3 \in G} Z(g_1, g_2, g_3)
\right),
\\
& = &
\frac{1}{|G|} \sum_{g_1,g_2,g_3 \in G} Z(g_1,g_2,g_3),
\\
& = &
Z_{T^3}\left( [X/G] \right),
\end{eqnarray}
where we have used the fact that the theta angle $\epsilon_R(+1) = 1$.
Thus, we see that the partition function of the $B^2K$-gauged theory
matches that of the orbifold $[X/G]$ for all $\epsilon_R$.
This is consistent with the original decomposition:  all universes
are identical, copies of $[X/G]$, so we see that, trivially,
for each $R \in \hat{K}$, we have recovered the corresponding universe
of the decomposition.

Next, we consider a less trivial case.
Specifically, consider the case $G = ({\mathbb Z}_2)^3$, $K = {\mathbb Z}_2$,
with extension
\begin{equation}
1 \: \longrightarrow \: B {\mathbb Z}_2 \: \longrightarrow \:
\tilde{\Gamma} \: \longrightarrow \: ({\mathbb Z}_2)^3 \: \longrightarrow \: 1
\end{equation}
of extension class $\omega_4 \in H^3(G,K)$, as discussed in
\cite[section 4.4]{Pantev:2022kpl}.
In this case, the decomposition is nontrivial:
\begin{equation}  \label{eq:decomp:3d:z23z2}
{\rm QFT}\left( [X/\tilde{\Gamma}] \right) \: = \:
{\rm QFT}\left( ]X/G] \right) \, \coprod \,
{\rm QFT}\left( [X/G]_{\rm d.t.} \right),
\end{equation}
where the second copy of $[X/G]$ has nontrivial discrete torsion.

The partition function of $[X/\tilde{\Gamma}]$ on $Y = T^3$ takes the form
\begin{eqnarray}
Z_{Y=T^3}\left( [X/\tilde{\Gamma}] \right) & = &
\frac{1}{|K|^2 |G|} \sum_{z_1, z_2, z_3 \in K} \sum_{g_1, g_2, g_3 \in G, \epsilon_Y = 1} Z(g_1, g_2, g_3).
\end{eqnarray}
In this case, the constraint $\epsilon_Y(P) = 1$ on $G$ bundles arising
as $\tilde{\Gamma}$ bundles is nontrivial for $Y = T^3$,
and as discussed in \cite[section 4.4]{Pantev:2022kpl},
\begin{eqnarray}
Z_{Y=T^3}\left( [X/\tilde{\Gamma}] \right) & = &
\frac{|K|}{|G|} \sum_{g_{1-3} \in G, \epsilon_Y = 1} Z(g_1,g_2,g_3),
\\
& = & 
Z_{T^3}\left( [X/G] \, \coprod \, [X/G]_{\rm d.t.} \right).
\end{eqnarray}

Next, we gauge the $B^2 K$ action on the theory above.
From the general prescription~(\ref{eq:gauged-b2k-2gp-orb-t3}),
\begin{eqnarray}
\lefteqn{
Z_{Y = T^3}\left( \left[ [X/\tilde{\Gamma}] / B^2 K\right] \right)
} \nonumber \\
& = &
\frac{1}{|K|} \sum_{\gamma \in H^3(T^3,K) = K}
\epsilon_R(\gamma)
\left( \frac{1}{|K|^2 |G|} \sum_{z_1, z_2, z_3 \in K} \sum_{g_1, g_2, g_3 \in G, \epsilon_Y = \gamma} Z(g_1, g_2, g_3)
\right),
\\
& = &
\frac{1}{|K|} \epsilon_R(+1) \left( 
\frac{|K|}{|G|} \sum_{g_{1-3} \in G, \epsilon_Y = +1}
Z(g_1, g_2, g_3) \right)
\nonumber \\
& & 
\: + \:
\frac{1}{|K|} \epsilon_R(-1) \left( 
\frac{|K|}{|G|} \sum_{g_{1-3} \in G, \epsilon_Y = -1}
Z(g_1, g_2, g_3) \right).
\end{eqnarray}

In the case that $\epsilon_R(-1) = +1$,
\begin{eqnarray}
Z_{Y = T^3}\left( \left[ [X/\tilde{\Gamma}] / B^2 K\right] \right)
& = &
\frac{1}{|G|} \sum_{g_1,g_2,g_3 \in G} Z(g_1,g_2,g_3),
\\
& = &
Z_{Y=T^3}\left( [X/G] \right),
\end{eqnarray}
consistent with
\begin{equation}
{\rm QFT}\left( \left[ [X/\tilde{\Gamma}] / B^2 K \right] \right)
\: = \:
{\rm QFT}\left( [X/G] \right),
\end{equation}
recovering one of the two universes of the decomposition~(\ref{eq:decomp:3d:z23z2}).

In the case that $\epsilon_R(-1) = -1$,
\begin{eqnarray}
Z_{Y = T^3}\left( \left[ [X/\tilde{\Gamma}] / B^2 K\right] \right)
& = &
\frac{1}{|G|} \sum_{g_1,g_2,g_3 \in G}
\epsilon_{Y}(g_1,g_2,g_3) Z(g_1,g_2,g_3),
\\
& = &
Z_{Y=T^3}\left( [X/G]_{\rm d.t.} \right),
\end{eqnarray}
where $\epsilon_Y(g_1,g_2,g_3)$ represents the phase arising from
discrete torsion in this context \cite{Dijkgraaf:1989pz,Sharpe:2000qt}
(which is a minus sign on the sectors which were excluded in the
original $\tilde{\Gamma}$ orbifold), consistent with
\begin{equation}
{\rm QFT}\left( \left[ [X/\tilde{\Gamma}] / B^2 K \right] \right)
\: = \: 
{\rm QFT}\left( [X/G]_{\rm d.t.} \right).
\end{equation}
In this case, we
recover the other universe of the
decomposition~(\ref{eq:decomp:3d:z23z2}), as expected.

\subsubsection{Defects}

In this section we will consider a three-dimensional
defect in a four-dimensional low-energy effective orbifold
$[X/\tilde{\Gamma}]$ by a 2-group $\tilde{\Gamma}$:
\begin{equation}
1 \: \longrightarrow \: BK \: \longrightarrow \:
\tilde{\Gamma} \: \longrightarrow \: G \: \longrightarrow \: 1,
\end{equation}
where $BK$ acts trivially,
and the extension is classified by $\omega \in H^3(G,K)$,
as in \cite{Pantev:2022kpl} and as reviewed in section~\ref{sect:3dorb}.

Because the $BK$ acts trivially, the resulting theory has a 
global two-form symmetry.  In a four-dimensional theory, this would not
result in a decomposition, but in a three-dimensional theory,
as along a defect $Y$, it does.

Restrict the four-dimensional theory above to a three-dimensional
submanifold $Y$ of spacetime, the location of the defect.  
The restriction of the four-dimensinoal
theory to $Y$ is a theory with a global two-form symmetry, and so
decomposes.  We will product an analogue of a condensation defect by gauging
that global two-form symmetry, as reviewed in 
section~\ref{sect:3dorb}, which selects out a universe (depending
upon the theta angle chosen).

So, for each three-dimensional submanifold $Y$, we now have a collection
of defects, one for each universe in the decomposition of a three-dimensional
orbifold $[X/\tilde{\Gamma}]$.

Now, let us consider fusion rules.  Following section~\ref{sect:3dorb},
the defect s obtained by gauging a 2-form symmetry $B^2 K$ on a theory
on the three-dimensional space $Y$, which means the path integral
\begin{itemize}
\item sums over $K$ 2-gerbes, and then,
\item for each $K$ 2-gerbe, sums over 2-gerbe twisted 
$\tilde{\Gamma}$-bundles and maps into $X$.
\end{itemize}
In principle, just as in the two-dimensional case, in the path integral
of the fusion of the two defects along the same submanifold $Y$, one would
like to tensor together the $K$ 2-gerbes and the twisted 
$\tilde{\Gamma}$-bundles, for which one runs into analogues of the
same issues encountered in two-dimensional examples previously.

First, let us discuss tensoring the 2-gerbes.  Our analysis here
is very similar to that in the two-dimensional orbifold examples
earlier, and we shall closely follow the same pattern.
Since $K$ is abelian, it suffices to assume that $K$ is cyclic.
Suppose one defect is defined by gauging $B^2 {\mathbb Z}_p$,
and the other by gauging $B^2 {\mathbb Z}_k$,
where both ${\mathbb Z}_p, {\mathbb Z}_k \subset K$.
Formally, the product of these 2-gerbes maps to a
${\mathbb Z}_{pk}$ 2-gerbe; however, that overcounts physical degrees
of freedom, as a common ${\mathbb Z}_{ {\rm gcd}(p,k)}$ 2-gerbe
can be eliminated through a change of variables.
Proceeding in a fashion closely analogous to the two-dimensional
case, the relation~(\ref{eq:gcd-lcm}) induces
\begin{equation}
H^3(Y, {\mathbb Z}_{ {\rm gcd}(p,k) } ) \: \longrightarrow \:
H^3( Y, {\mathbb Z}_p\times {\mathbb Z}_k) \:
\longrightarrow \: H^3( Y, {\mathbb Z}_{ {\rm lcm}(p,k) } )
\: \longrightarrow \: 0.
\end{equation}
Put more simply, this means that the product of
${\mathbb Z}_p$ and
${\mathbb Z}_k$ 2-gerbes can be described as
${\mathbb Z}_{ {\rm lcm}(p,k) } \subset K$ 2-gerbes, and the mapping to
${\mathbb Z}_{ {\rm lcm}(p,k) }$ 2-gerbes has, as fiber,
${\mathbb Z}_{ {\rm gcd}(p,k) }$ 2-gerbes.

Next, we turn to the $\tilde{\Gamma}$ bundles.
As in the case of two-dimensional orbifolds, we do not know of
a way to simply tensor together the bundles in general.  However,
as in our previous discussion, we can instead borrow a trick from
OPE computations of anomalies, and compute the fusion products by
replacing $Y$ with a box $Y \times I$, with the defects at either boundary.
As before, since $I$ is contractible,
the path integral sums over isomorphisms between the data
at each boundary.

Assembling these pieces, and using results for partition functions for
$\tilde{\Gamma}$ orbifolds and $B^2K$ orbifolds thereof,
we find that in a large mass limit, the partition function of the
fusion product of one defect obtained by gauging $B^2 {\mathbb Z}_p$
and another obtained by gauging $B^2 {\mathbb Z}_k$ on
$Y = T^3$ is
\begin{equation}
\frac{ {\rm gcd}(p,k) }{ | {\mathbb Z}_{ {\rm lcm}(p,k) } | }
\sum_{k \in H^3(Y, {\mathbb Z}_{\rm lcm})}
\left[ 
\frac{ |K|^2 }{ |G|^2 } \sum_{g_{1-3} \in G, \epsilon_Y = k}
\sum_{h_{1-3} \in G, \epsilon_Y = k} \sum_{\gamma \in G}
\epsilon_{\ell_1}(k) \,
\epsilon_{\ell_2}(k) \, Z(g_{1-3}, h_{1-3}, k)
\right],
\end{equation}
where
\begin{equation}
g_i g_j \: = \: g_j g_i, \: \: \:
h_i h_j \: = \: h_j h_i, \: \: \:
g_i \: = \: \gamma h_i \gamma^{-1},
\end{equation}
and 
\begin{equation}
\epsilon_Y(g_1, g_2, g_3) \: = \: k \: = \: \epsilon_Y(h_1, h_2, h_3).
\end{equation}

Suppose for example that $G$ is abelian, then $\gamma$ effectively
decouples as $g_i = h_i$, and the partition function above reduces to
\begin{eqnarray}
\lefteqn{
\frac{ {\rm gcd}(p,k) }{ | {\mathbb Z}_{ {\rm lcm}(p,k) } | }
\sum_{k \in H^3(Y, {\mathbb Z}_{\rm lcm})}
\left[ \frac{ |K|^2 }{ |G|^2 } |G| \sum_{g_{1-3} \in G, \epsilon_Y = k}
\epsilon_{\ell_1 + \ell_2}(k) \, Z(g_{1-3}, k)
\right],
} \nonumber \\
& = &
\left( {\rm gcd}(p,k) \right) |K| Z_Y\left( \left[
[X/\tilde{\Gamma}]|_Y / B^2 {\mathbb Z}_{ {\rm lcm}(p,k) } \right]
\right).
\end{eqnarray}

For example, if $G = ({\mathbb Z}_2)^3$, $K = {\mathbb Z}_2$,
and $p = k = 2$, then denoting the $\ell$th defect by $S_{\ell}(Y)$,
and assuming we did not drop any factors,
this becomes
\begin{eqnarray}
S_{\ell_1}(T^3) \times S_{\ell_2}(T^3) & = &
(2) | {\mathbb Z}_2 | \, S_{\ell_1 + \ell_2 \mod 2}(T^3),
\\
& = &
(4)  S_{\ell_1 + \ell_2 \mod 2}(T^3).
\end{eqnarray}

\section*{Acknowledgements}

We would like to thank Y.~Choi, S.~Gukov, H.-T.~Lam, T.~Pantev, S.-H.~Shao,
and M.~Yu for useful discussions, and especially
T.~Vandermeulen for both numerous discussions and initial collaboration.
L.L.~and D.R.~further thank the Simons Center for Geometry and Physics for hospitality during 2022 Summer Workshop, at which parts of this work was carried out.
D.R.~was partially supported by
NSF grant PHY-1820867.
E.S.~was partially supported by NSF grant
PHY-2014086.


\begin{thebibliography}{199}

\addcontentsline{toc}{section}{References}

\bibitem{Hellerman:2006zs}
S.~Hellerman, A.~Henriques, T.~Pantev, E.~Sharpe and M.~Ando, ``Cluster
  decomposition, T-duality, and gerby CFT's,''
  \href{https://doi.org/10.4310/ATMP.2007.v11.n5.a2}{\emph{Adv. Theor. Math.
  Phys.} {\bfseries 11} (2007) 751--818},
  [\href{https://arxiv.org/abs/hep-th/0606034}{{\ttfamily hep-th/0606034}}].

\bibitem{Sharpe:2022ene}
E.~Sharpe, ``An introduction to decomposition,''
  [\href{https://arxiv.org/abs/2204.09117}{{\ttfamily arXiv:2204.09117}}].


\bibitem{Pantev:2005rh}
T.~Pantev and E.~Sharpe, ``Notes on gauging noneffective group actions,''
  [\href{https://arxiv.org/abs/hep-th/0502027}{{\ttfamily hep-th/0502027}}].
  
\bibitem{Pantev:2005wj}
T.~Pantev and E.~Sharpe, ``String compactifications on Calabi-Yau
  stacks,'' \href{https://doi.org/10.1016/j.nuclphysb.2005.10.035}{\emph{Nucl.
  Phys. B} {\bfseries 733} (2006) 233--296},
  [\href{https://arxiv.org/abs/hep-th/0502044}{{\ttfamily hep-th/0502044}}].

\bibitem{Pantev:2005zs}
T.~Pantev and E.~Sharpe, ``GLSM's for gerbes (and other toric stacks),''
  \href{https://doi.org/10.4310/ATMP.2006.v10.n1.a4}{\emph{Adv. Theor. Math.
  Phys.} {\bfseries 10} (2006) 77--121},
  [\href{https://arxiv.org/abs/hep-th/0502053}{{\ttfamily hep-th/0502053}}].




\bibitem{Durhuus:1993cq}
B.~Durhuus and T.~Jonsson, ``Classification and construction of unitary
  topological field theories in two-dimensions,''
  \href{https://doi.org/10.1063/1.530752}{\emph{J. Math. Phys.} {\bfseries 35}
  (1994) 5306--5313}, [\href{https://arxiv.org/abs/hep-th/9308043}{{\ttfamily
  hep-th/9308043}}].

\bibitem{Moore:2006dw}
G.~W. Moore and G.~Segal, ``D-branes and K-theory in 2d topological field
  theory,''  [\href{https://arxiv.org/abs/hep-th/0609042}{{\ttfamily
  hep-th/0609042}}].

\bibitem{Komargodski:2020mxz}
Z.~Komargodski, K.~Ohmori, K.~Roumpedakis and S.~Seifnashri, ``Symmetries
  and strings of adjoint QCD$_{2}$,''
  \href{https://doi.org/10.1007/JHEP03(2021)103}{\emph{JHEP} {\bfseries 03}
  (2021) 103}, [\href{https://arxiv.org/abs/2008.07567}{{\ttfamily
  arXiv:2008.07567}}].

\bibitem{Huang:2021zvu}
T.-C. Huang, Y.-H. Lin and S.~Seifnashri, ``Construction of
  two-dimensional topological field theories with non-invertible symmetries,''
  \href{https://doi.org/10.1007/JHEP12(2021)028}{\emph{JHEP} {\bfseries 12}
  (2021) 028}, [\href{https://arxiv.org/abs/2110.02958}{{\ttfamily
  arXiv:2110.02958}}].

\bibitem{Roumpedakis:2022aik}
K.~Roumpedakis, S.~Seifnashri and S.-H. Shao, ``Higher gauging and
  non-invertible condensation defects,''
  [\href{https://arxiv.org/abs/2204.02407}{{\ttfamily arXiv:2204.02407}}].

\bibitem{Choi:2022zal}
Y.~Choi, C.~C\'ordova, P.-S. Hsin, H.~T. Lam and S.-H. Shao,
  ``Non-invertible condensation, duality, and triality defects in 3+1
  dimensions,''  [\href{https://arxiv.org/abs/2204.09025}{{\ttfamily
  arXiv:2204.09025}}].

\bibitem{Kong:2014qka}
L.~Kong and X.-G. Wen, ``Braided fusion categories, gravitational
  anomalies, and the mathematical framework for topological orders in any
  dimensions,''  [\href{https://arxiv.org/abs/1405.5858}{{\ttfamily arXiv:1405.5858}}].

\bibitem{Else:2017yqj}
D.~V. Else and C.~Nayak, ``Cheshire charge in (3+1)-dimensional
  topological phases,''
  \href{https://doi.org/10.1103/PhysRevB.96.045136}{\emph{Phys. Rev. B}
  {\bfseries 96} (2017) 045136},
  [\href{https://arxiv.org/abs/1702.02148}{{\ttfamily arXiv:1702.02148}}].

\bibitem{Carqueville:2017ono}
N.~Carqueville, I.~Runkel and G.~Schaumann, ``Line and surface defects in
  Reshetikhin-Turaev TQFT,''
  [\href{https://arxiv.org/abs/1710.10214}{{\ttfamily arXiv:1710.10214}}].

\bibitem{Carqueville:2018sld}
N.~Carqueville, I.~Runkel and G.~Schaumann, ``Orbifolds of
  Reshetikhin-Turaev TQFTs,'' {\emph{Theor. Appl. Categor.} {\bfseries 35}
  (2020) 513--561}, [\href{https://arxiv.org/abs/1809.01483}{{\ttfamily
  arXiv:1809.01483}}].

\bibitem{Hsin:2019fhf}
P.-S. Hsin and A.~Turzillo, ``Symmetry-enriched quantum spin liquids in (3
  + 1)$d$,'' \href{https://doi.org/10.1007/JHEP09(2020)022}{\emph{JHEP}
  {\bfseries 09} (2020) 022},
  [\href{https://arxiv.org/abs/1904.11550}{{\ttfamily arXiv:1904.11550}}].

\bibitem{Gaiotto:2019xmp}
D.~Gaiotto and T.~Johnson-Freyd, ``Condensations in higher categories,''
  [\href{https://arxiv.org/abs/1905.09566}{{\ttfamily arXiv:1905.09566}}].

\bibitem{Mulevicius:2020bat}
V.~Mulevi\v{c}ius and I.~Runkel, ``Constructing modular categories from
  orbifold data,''  [\href{https://arxiv.org/abs/2002.00663}{{\ttfamily
  arXiv:2002.00663}}].

\bibitem{Johnson-Freyd:2020usu}
T.~Johnson-Freyd, ``On the classification of topological orders,''
  \href{https://doi.org/10.1007/s00220-022-04380-3}{\emph{Commun. Math. Phys.}
  {\bfseries 393} (2022) 989--1033},
  [\href{https://arxiv.org/abs/2003.06663}{{\ttfamily arXiv:2003.06663}}].

\bibitem{Kong:2020cie}
L.~Kong, T.~Lan, X.-G. Wen, Z.-H. Zhang and H.~Zheng, ``Algebraic higher
  symmetry and categorical symmetry -- a holographic and entanglement view of
  symmetry,''
  \href{https://doi.org/10.1103/PhysRevResearch.2.043086}{\emph{Phys. Rev.
  Res.} {\bfseries 2} (2020) 043086},
  [\href{https://arxiv.org/abs/2005.14178}{{\ttfamily arXiv:2005.14178}}].

\bibitem{Kong:2020wmn}
L.~Kong, Y.~Tian and Z.-H. Zhang, ``Defects in the 3-dimensional toric
  code model form a braided fusion 2-category,''
  \href{https://doi.org/10.1007/JHEP12(2020)078}{\emph{JHEP} {\bfseries 12}
  (2020) 078}, [\href{https://arxiv.org/abs/2009.06564}{{\ttfamily
  arXiv:2009.06564}}].

\bibitem{Johnson-Freyd:2020twl}
T.~Johnson-Freyd, ``(3+1)d topological orders with only a
  $\mathbb{Z}_2$-charged particle,''
  [\href{https://arxiv.org/abs/2011.11165}{{\ttfamily arXiv:2011.11165}}].

\bibitem{Carqueville:2021dbv}
N.~Carqueville, V.~Mulevicius, I.~Runkel, G.~Schaumann and D.~Scherl,
  ``Orbifold graph TQFTs,''
  [\href{https://arxiv.org/abs/2101.02482}{{\ttfamily arXiv:2101.02482}}].

\bibitem{Koppen:2021kry}
V.~Koppen, V.~Mulevicius, I.~Runkel and C.~Schweigert, ``Domain walls
  between 3d phases of Reshetikhin-Turaev TQFTs,''
  [\href{https://arxiv.org/abs/2105.04613}{{\ttfamily arXiv:2105.04613}}].

\bibitem{Carqueville:2021edn}
N.~Carqueville, V.~Mulevicius, I.~Runkel, G.~Schaumann and D.~Scherl,
  ``Reshetikhin-Turaev TQFTs close under generalised orbifolds,''
  [\href{https://arxiv.org/abs/2109.04754}{{\ttfamily arXiv:2109.04754}}].

\bibitem{Heidenreich:2021xpr}
B.~Heidenreich, J.~McNamara, M.~Montero, M.~Reece, T.~Rudelius and
  I.~Valenzuela, ``Non-invertible global symmetries and completeness of
  the spectrum,'' \href{https://doi.org/10.1007/JHEP09(2021)203}{\emph{JHEP}
  {\bfseries 09} (2021) 203},
  [\href{https://arxiv.org/abs/2104.07036}{{\ttfamily arXiv:2104.07036}}].

\bibitem{Nguyen:2021yld}
M.~Nguyen, Y.~Tanizaki and M.~\"Unsal, ``Semi-Abelian gauge theories,
  non-invertible symmetries, and string tensions beyond $N$-ality,''
  \href{https://doi.org/10.1007/JHEP03(2021)238}{\emph{JHEP} {\bfseries 03}
  (2021) 238}, [\href{https://arxiv.org/abs/2101.02227}{{\ttfamily
  arXiv:2101.02227}}].

\bibitem{Nguyen:2021naa}
M.~Nguyen, Y.~Tanizaki and M.~\"Unsal,
``Noninvertible 1-form symmetry and Casimir scaling in 2d Yang-Mills theory,''
\href{https://journals.aps.org/prd/abstract/10.1103/PhysRevD.104.065003}{\emph{Phys. Rev. D} \textbf{104} (2021) 065003},
[\href{https://arxiv.org/abs/2104.01824}{{\tt arXiv:2104.01824}}].


\bibitem{Sharpe:2021srf}
E.~Sharpe,
``Topological operators, noninvertible symmetries and decomposition,''
[\href{https://arxiv.org/abs/2108.13423}{{\tt arXiv:2108.13423}}].

\bibitem{Koide:2021zxj}
M.~Koide, Y.~Nagoya and S.~Yamaguchi, ``Non-invertible topological defects
  in 4-dimensional $\mathbb {Z}_2$ pure lattice gauge theory,''
  \href{https://doi.org/10.1093/ptep/ptab145}{\emph{PTEP} {\bfseries 2022}
  (2022) 013B03}, [\href{https://arxiv.org/abs/2109.05992}{{\ttfamily
  arXiv:2109.05992}}].

\bibitem{Choi:2021kmx}
Y.~Choi, C.~C\'ordova, P.-S. Hsin, H.~T. Lam and S.-H. Shao, ``Noninvertible
  duality defects in 3+1 dimensions,''
  \href{https://doi.org/10.1103/PhysRevD.105.125016}{\emph{Phys. Rev. D}
  {\bfseries 105} (2022) 125016},
  [\href{https://arxiv.org/abs/2111.01139}{{\ttfamily arXiv:2111.01139}}].

\bibitem{Kaidi:2021xfk}
J.~Kaidi, K.~Ohmori and Y.~Zheng, ``Kramers-Wannier-like duality defects
  in (3+1)d gauge theories,''
  \href{https://doi.org/10.1103/PhysRevLett.128.111601}{\emph{Phys. Rev. Lett.}
  {\bfseries 128} (2022) 111601},
  [\href{https://arxiv.org/abs/2111.01141}{{\ttfamily arXiv:2111.01141}}].

\bibitem{Bhardwaj:2022yxj}
L.~Bhardwaj, L.~Bottini, S.~Schafer-Nameki and A.~Tiwari, ``Non-invertible
  higher-categorical symmetries,''
  [\href{https://arxiv.org/abs/2204.06564}{{\ttfamily arXiv:2204.06564}}].

\bibitem{Hayashi:2022fkw}
Y.~Hayashi and Y.~Tanizaki, ``Non-invertible self-duality defects of
  Cardy-Rabinovici model and mixed gravitational anomaly,''
  [\href{https://arxiv.org/abs/2204.07440}{{\ttfamily arXiv:2204.07440}}].

\bibitem{Arias-Tamargo:2022nlf}
G.~Arias-Tamargo and D.~Rodriguez-Gomez, ``Non-invertible symmetries from
  discrete gauging and completeness of the spectrum,''
  [\href{https://arxiv.org/abs/2204.07523}{{\ttfamily arXiv:2204.07523}}].

\bibitem{Choi:2022jqy}
Y.~Choi, H.~T. Lam and S.-H. Shao, ``Non-invertible global symmetries in
  the Standard Model,''  [\href{https://arxiv.org/abs/2205.05086}{{\ttfamily
  arXiv:2205.05086}}].

\bibitem{Cordova:2022ieu}
C.~C\'ordova and K.~Ohmori, ``Non-invertible chiral symmetry and exponential
  hierarchies,''  [\href{https://arxiv.org/abs/2205.06243}{{\ttfamily
  arXiv:2205.06243}}].

\bibitem{Antinucci:2022eat}
A.~Antinucci, G.~Galati and G.~Rizi,
``On continuous 2-category symmetries and Yang-Mills theory,''
[\href{https://arxiv.org/abs/2206.05646}{\tt arXiv:2206.05646}].


\bibitem{Bashmakov:2022jtl}
V.~Bashmakov, M.~Del~Zotto and A.~Hasan, ``On the 6d origin of
  non-invertible symmetries in 4d,''
  [\href{https://arxiv.org/abs/2206.07073}{{\ttfamily arXiv:2206.07073}}].

\bibitem{Damia:2022bcd}
J.~A. Damia, R.~Argurio and E.~Garcia-Valdecasas, ``Non-invertible defects
  in 5d, boundaries and holography,''
  [\href{https://arxiv.org/abs/2207.02831}{{\ttfamily arXiv:2207.02831}}].

\bibitem{Choi:2022rfe}
Y.~Choi, H.~T.~Lam and S.~H.~Shao,
``Non-invertible time-reversal symmetry,''
[\href{https://arxiv.org/abs/2208.04331}{\tt arXiv:2208.04331}].


\bibitem{Sharpe:2019ddn}
E.~Sharpe, ``Undoing decomposition,''
  \href{https://doi.org/10.1142/S0217751X19502336}{\emph{Int. J. Mod. Phys. A}
  {\bfseries 34} (2020) 1950233},
  [\href{https://arxiv.org/abs/1911.05080}{{\ttfamily arXiv:1911.05080}}].


\bibitem{Pantev:2022pbf}
T.~Pantev and E.~Sharpe, ``Decomposition in Chern-Simons theories in three
  dimensions,''  [\href{https://arxiv.org/abs/2206.14824}{{\ttfamily
  arXiv:2206.14824}}].




\bibitem{Hsin:2020nts}
P.-S. Hsin and H.~T. Lam, ``Discrete theta angles, symmetries and
  anomalies,''
  \href{https://doi.org/10.21468/SciPostPhys.10.2.032}{\emph{SciPost Phys.}
  {\bfseries 10} (2021) 032},
  [\href{https://arxiv.org/abs/2007.05915}{{\ttfamily arXiv:2007.05915}}].


\bibitem{Bhardwaj:2022lsg}
L.~Bhardwaj, S.~Schafer-Nameki and J.~Wu,
``Universal non-invertible symmetries,''
[\href{https://arxiv.org/abs/2208.05973}{\tt arXiv:2208.05973}].

\bibitem{Bartsch:2022mpm}
T.~Bartsch, M.~Bullimore, A.~E.~V.~Ferrari and J.~Pearson,
``Non-invertible symmetries and higher representation theory I,''
[\href{https://arxiv.org/abs/2208.05993}{\tt arXiv:2208.05993}].




\bibitem{Robbins:2020msp}
D.~Robbins, E.~Sharpe and T.~Vandermeulen, ``A generalization of
  decomposition in orbifolds,''
  \href{https://doi.org/10.1007/JHEP10(2021)134}{\emph{JHEP} {\bfseries 21}
  (2020) 134}, [\href{https://arxiv.org/abs/2101.11619}{{\ttfamily
  arXiv:2101.11619}}].




\bibitem{Pantev:2022kpl}
T.~Pantev, D.~Robbins, E.~Sharpe and T.~Vandermeulen, ``Orbifolds by
  2-groups and decomposition,''
  [\href{https://arxiv.org/abs/2204.13708}{{\ttfamily arXiv:2204.13708}}].


\bibitem{rsvtoappear}
D.~G. Robbins, E.~Sharpe and T.~Vandermeulen, to appear.

\bibitem{Ginsparg:1988ui}
P.~H. Ginsparg, ``Applied conformal field theory,''  pp. 1-168 in \emph{{Les Houches
  1988 Summer School in Theoretical Physics: Fields, Strings, Critical Phenomena}},
  eds. E. Br\'ezin, J. Zinn-Justin, Elsevier, 1989, [\href{https://arxiv.org/abs/hep-th/9108028}{{\ttfamily
  hep-th/9108028}}].



\bibitem{Dijkgraaf:1989pz}
R.~Dijkgraaf and E.~Witten, ``Topological gauge theories and group
  cohomology,'' \href{https://doi.org/10.1007/BF02096988}{\emph{Commun. Math.
  Phys.} {\bfseries 129} (1990) 393-429}.
  
\bibitem{Sharpe:2000qt}
E.~R. Sharpe, ``Analogues of discrete torsion for the M theory three
  form,'' \href{https://doi.org/10.1103/PhysRevD.68.126004}{\emph{Phys. Rev. D}
  {\bfseries 68} (2003) 126004},
  [\href{https://arxiv.org/abs/hep-th/0008170}{{\ttfamily hep-th/0008170}}].




\bibitem{Sharpe:2003cs}
E.~Sharpe,
``Discrete torsion and shift orbifolds,''
\href{https://www.sciencedirect.com/science/article/pii/S0550321303004127}{Nucl. Phys. B \textbf{664} (2003) 21-44},
[\href{https://arxiv.org/abs/hep-th/0302152}{\tt hep-th/0302152}].

\bibitem{Cheng:2022nso}
P.~Cheng, I.~V.~Melnikov and R.~Minasian,
``Flat equivariant gerbes: holonomies and dualities,''
\href{https://arxiv.org/abs/2207.06885}{\tt arXiv:2207.06885}.

\bibitem{clement} A. Cl\'ement, {\it Integral cohomology of finite
Postnikov towers}, Ph.D. thesis, Universit\'e de Lausanne, 2002,
available at
\href{http://doc.rero.ch/record/482/files/Clement_these.pdf}{http://doc.rero.ch/record/482/files/Clement\_these.pdf}.

\bibitem{stackexchange} \href{https://math.stackexchange.com/questions/43142/maps-between-eilenberg-maclane-spaces}{https://math.stackexchange.com/questions/43142/maps-between-eilenberg-maclane-spaces}


\bibitem{Hellerman:2010fv}
S.~Hellerman and E.~Sharpe, ``Sums over topological sectors and
  quantization of Fayet-Iliopoulos parameters,''
  \href{https://doi.org/10.4310/ATMP.2011.v15.n4.a7}{\emph{Adv. Theor. Math.
  Phys.} {\bfseries 15} (2011) 1141--1199},
  [\href{https://arxiv.org/abs/1012.5999}{{\ttfamily arXiv:1012.5999}}].

\bibitem{Blau:1993hj}
M.~Blau and G.~Thompson, ``Lectures on 2-d gauge theories: Topological
  aspects and path integral techniques,''  in \emph{{Summer School in
  High-energy Physics and Cosmology (Includes Workshop on Strings, Gravity, and
  Related Topics 29-30 Jul 1993)}}, pp.~175--244, 
  [\href{https://arxiv.org/abs/hep-th/9310144}{{\ttfamily hep-th/9310144}}].

\bibitem{Verlinde:1988sn}
E.~P. Verlinde, ``Fusion rules and modular transformations in 2d conformal
  field theory,''
  \href{https://doi.org/10.1016/0550-3213(88)90603-7}{\emph{Nucl. Phys. B}
  {\bfseries 300} (1988) 360--376}.

\bibitem{gukovprivate}
S.~Gukov, private communication.



\bibitem{Kapustin:2010if}
A.~Kapustin and N.~Saulina,
``Surface operators in 3d topological field theory and 2d rational conformal field theory,''
[\href{https://arxiv.org/abs/1012.0911}{\tt arXiv:1012.0911}].

\bibitem{Kapustin:2010hk}
A.~Kapustin and N.~Saulina,
``Topological boundary conditions in abelian Chern-Simons theory,''
\href{https://doi.org/10.1016/j.nuclphysb.2010.12.017
}{\emph{Nucl. Phys. B} \textbf{845} (2011) 393-435},
[\href{https://arxiv.org/abs/1008.0654}{\tt arXiv:1008.0654}].





\bibitem{Peskin:1995ev}
M.~E. Peskin and D.~V. Schroeder, \emph{{An introduction to quantum field
  theory}}.
\newblock Addison-Wesley, Reading, USA, 1995.



\end{thebibliography}
\end{document}